\documentclass{amsart} 
\usepackage{amssymb, amsmath}
\usepackage{amsfonts}
\usepackage{float}
\usepackage{graphics,color}
\usepackage{graphicx}
\usepackage{textcomp}
\usepackage{verbatim}
\usepackage{hyperref}

\def\bt{\begin{thm}}
\def\et{\end{thm}}
\def\bl{\begin{lem}}
\def\el{\end{lem}}
\def\bd{\begin{defi}}
\def\ed{\end{defi}}
\def\bc{\begin{cor}}
\def\ec{\end{cor}}
\def\bp{\begin{proof}}
\def\ep{\end{proof}}
\def\br{\begin{rem}}
\def\er{\end{rem}}

\def\Forall{\text{ } \forall \:}
\def\d{\, \mathrm{d}}

\def\be{\begin{equation}}
\def\ee{\end{equation}}
\def\bes{\begin{equation*}}
\def\ees{\end{equation*}}
\def\bea{\begin{equation} \begin{aligned}}
\def\eea{\end{aligned} \end{equation}}
\def\beas{\begin{equation*} \begin{aligned}}
\def\eeas{\end{aligned} \end{equation*}}

\newtheorem{thm}{Theorem}[section]
\newtheorem{lem}{Lemma}[section]
\newtheorem{defi}{Definition}[section]

\newtheorem{prop}[thm]{Proposition}

\newtheorem{rem}{Remark}[section]
\newtheorem{cor}{Corollary}[section]

\newcommand{\fraction}[2]{\frac{\textstyle #1}{\textstyle #2}}
\newcommand{\red}[1]{\textcolor{red}{#1}}
\newcommand{\blue}[1]{\textcolor{blue}{#1}}

\title[Cahn-Hilliard equation with Onsager mobility]{Dynamic Transitions for Quasilinear Systems and Cahn-Hilliard equation with Onsager mobility}
\author[Liu]{Honghu Liu}
\address[HL]{Department of Mathematics,
Indiana University, Bloomington, IN 47405} \email{liu40@indiana.edu}

\author[Sengul]{Taylan Sengul}
\address[TS]{Department of Mathematics,
Indiana University, Bloomington, IN 47405}
\email{msengul@indiana.edu}

\author[Wang]{Shouhong Wang}
\address[SW]{Department of Mathematics,
Indiana University, Bloomington, IN 47405}
\email{showang@indiana.edu, http://www.indiana.edu/\texttildelow
fluid}

\thanks{The authors are grateful to an anonymous referee for his/her very detailed and insightful comments, which substantially improved the paper. The work  was supported in part by the
Office of Naval Research and by the National Science Foundation.}


\date{\today}

\begin{document}

\begin{abstract}
The main objectives of this article  are two-fold. First, we study
the effect of the nonlinear Onsager mobility on the phase transition
and on the well-posedness of the Cahn-Hilliard equation modeling a
binary system. It is shown in particular  that the dynamic
transition is essentially independent of the nonlinearity of the
Onsager mobility. However, the nonlinearity of the mobility does
cause substantial technical difficulty for the well-posedness and
for carrying out the dynamic transition analysis. For this reason,
as a second objective, we introduce a systematic approach to deal
with phase transition problems modeled by quasilinear partial
differential equation, following the ideas of the dynamic transition
theory developed in Ma and Wang \cite{MW05, MW}.
\end{abstract}

\maketitle
\tableofcontents
\section{Introduction}
The Cahn-Hilliard equation is a basic model in material science, as
it characterizes important qualitative features of binary systems.
The model has been intensively studied, especially in the case of
constant mobility; see among many others \cite{BB99, EG96, Ell89,
MW09a, MW09b, MZ05, NS84}. However,  the dependence of the
mobility on the concentration is very much relevant for physical
applications, and a concentration dependent mobility appeared in the
original derivation of the Cahn-Hilliard equation in \cite{CH58}. In
this case, the modeling equation is no longer a  semilinear equation,
and become  a quasilinear
equation,  which makes the problem much more challenging.

The main objectives of this article are to study the effect of  the
nonlinearity of the Onsager mobility on the phase transition
dynamics and on the well-posedness of the model, and to introduce a
systematic approach for studying phase transitions for such
quasilinear systems.

First, for a quasilinear dynamical system as the Cahn-Hilliard
equation with the Onsager mobility, the main difficulty comes from
the regularity loss through the nonlinear terms involving the
highest order spatial derivatives. This has to be compensated by the
regularizing properties of the linear operator. In particular, the
so called maximal regularity property \cite{DaLun88, Simonett95} is
essential to guarantee the existence of a center manifold for a
quasilinear system. This can be achieved by working in more regular
function spaces \cite{bates98, DaLun88, Lun95, Mie91, Simonett95};
see Section \ref{app. CM for quasilinear} for more details. Under
this setup, we are able to derive the same approximation formulas
for center manifold functions for quasilinear systems as in
\cite{MW}. With these approximations at our disposal, the main ideas
and methods in the dynamic transition theory can then be applied to
studying quasilinear systems.

Second, by putting the  Cahn-Hilliard equation with Onsager mobility
in the framework just mentioned, we are able to derive the detailed
transition dynamics as for the constant mobility case, leading to
precise information on the type and structure of dynamic transition.
In particular, we derive that as for  the steady state bifurcation
case given by Hsia \cite{Hsi10},  the type of transition, the
critical temperature and the strength of deviation of solutions from
the homogenous state are all independent of the choices of the
nonlinearity of  the Onsager mobility.

Third, to set up the problem so that  we can use the center manifold
theory and  the approximation formulas for the center manifold
functions for quasilinear systems, we need to examine carefully the
well-posedness of the model.  In the constant mobility case, the
equation being semilinear, the well-posedness can be dealt with
using standard procedure for semilinear equations (see e.g.
\cite{Hen81}). However the well-posedness is an issue in the
non-constant mobility case and the results in this case are far from
being satisfactory. For  the two-dimensional case, the existence and
uniqueness of a classical solution has been established recently in
\cite{LQY06}. But for the three-dimensional case, we are not aware
of any such result except some partial results; see also \cite{BB99,
EG96, Sch07}. Hence we derive the existence and uniqueness theorems
of global strong solution with small initial data to the equation,
which is sufficient for the purposes of this paper.

This article is organized as follows: The model is presented in
Section 2, and the phase transitions for the model in a rectangular
domain is given in Section 3. Section 4 addresses the general
framework for dynamic transitions for quasilinear systems. Section 5
is devoted to the proofs of the phase transition results based on
the dynamic transition theory. The existence and uniqueness of
global strong solutions is analyzed in Section 6.

\section{The Model} \label{section the model}
Consider a binary system consisting of elements A and B with molar
fractions $u_1$ and $1-u_1$, respectively. The free energy of the system is given
by
$$\mathcal{G}(u_1)=\int_\Omega \left(\frac{\alpha}{2} |\nabla u_1|^2+\Psi(u_1)\right) \, \mathrm{d} x,$$
where $\Omega$ is an open subset in $\mathbb{R}^3$ with  Lipschitz boundary $\partial \Omega$, $\alpha>0$ is a constant, and $\Psi(u_1)$, the homogeneous
free energy for a mean field model of binary systems at a fixed
temperature, is in the Hildebrand form:
$$\Psi(u_1)=RT\left( u_1 \ln u_1+(1-u_1)\ln(1-u_1) \right)+\gamma u_1 (1-u_1).$$
Here $R$ is the molar gas constant,  $T$ is the temperature of the system measured in Kelvin, and $\gamma>0$ is the coefficient of repulsive interaction between A and B.

The Cahn-Hilliard equation associated with the above free energy is the following; see \cite{CH58, NS84,
MW09a, Hsi10}:
\begin{equation} \label{CHE introduction}
\begin{split}
\frac{\partial u_1 }{\partial t} =& - \nabla \cdot J, \\
J=& -H(u_1) \nabla \mu \text{ and } \mu = \frac{\delta
\mathcal{G}}{\delta u_1}= -\alpha \Delta u_1 + \Psi'(u_1),
\end{split}
\end{equation}
where $J$ is the flux of type-A molecules, $H(u_1)$, a strictly positive function, is the Onsager
mobility measuring the strength of diffusion, $\mu$ is the
generalized chemical potential, and $\delta \mathcal{G}/ \delta u_1$
is the variational derivative of $\mathcal{G}$.

The above equation is supplemented with no-flux and Neumann boundary
conditions:
\begin{equation*}
\begin{split}
J \cdot \nu |_{\partial \Omega} =& 0, \\
\nabla u_1 \cdot \nu |_{\partial \Omega} =& 0,
\end{split}
\end{equation*}
which is equivalent to
\begin{equation}  \label{BC}
\frac{\partial u_1}{\partial \nu}|_{\partial \Omega}=0, \qquad
\frac{\partial \Delta u_1}{\partial \nu}|_{\partial \Omega}=0,
\end{equation}
where $\nu$ is the outward unit normal vector at the boundary
$\partial \Omega$. As a consequence of the no-flux boundary
condition, the mass is conserved:
\begin{equation} \label{mass conservation}
\frac{\mathrm{d}}{\mathrm{d} t} \int_\Omega u_1 \, \mathrm{d} x =0.
\end{equation}

Now representing the deviation of concentration around a homogenous
state $\overline{u}_1$ by $u=u_1-\overline{u}_1$ and approximating
$H(u_1)$ and $\Psi'(u_1)$ by their Taylor expansions about
$\overline{u}_1$, the equation governing the evolution of $u$ can be
stated as follows; see Hsia \cite{Hsi10}:
\begin{equation}
\begin{split}
\frac{\partial u}{\partial t}=& -H(\overline{u}_1)\Delta\left[\alpha\Delta u-b_1 u-b_2 u^2-b_3 u^3+o(u^3)\right] \\
&-H'(\overline{u}_1)\nabla\left[u\nabla(\alpha \Delta u-b_1 u-b_2 u^2+o(u^2))\right] \\
&-\frac{1}{2}H''(\overline{u}_1)\nabla\left[u^2\nabla(\alpha\Delta
u-b_1 u+o(u))\right].
\end{split}
\label{model}
\end{equation}
Here $H(\overline{u}_1)>0$ is the Onsager coefficient evaluated at
$u=\overline{u}_1$, and
\begin{equation*}
\begin{aligned}
& b_1=\frac{RT}{\overline{u}_1(1-\overline{u}_1)}-2\gamma, \\
& b_2=\frac{1}{2}RT \left(\frac{1}{(1-\overline{u}_1)^2}-\frac{1}{\overline{u}_1^2}\right),\\
& b_3=\frac{1}{3}RT
\left(\frac{1}{(1-\overline{u}_1)^3}+\frac{1}{\overline{u}_1^3}\right).
\end{aligned}
\end{equation*}
The boundary conditions in \eqref{BC} read:
\begin{equation} \label{BC2}
\frac{\partial u}{\partial \nu}|_{\partial \Omega}=0, \qquad
\frac{\partial \Delta u}{\partial \nu}|_{\partial \Omega}=0.
\end{equation}
Equation \eqref{model} is also supplemented with the following initial condition
\be \label{initial}
u(0) = \phi.
\ee
Due to the mass conservation \eqref{mass conservation}, we
assume in additional that
\begin{equation} \label{meanzero}
\int_\Omega u\, \mathrm{d}x=0.
\end{equation}

\section{Effects of the Onsager Mobility on Phase Transition Dynamics}  \label{section Main thm}
In this section we present our theorems describing the phase
transitions of Cahn-Hilliard equation in a rectangular box $\Omega =
\prod_{i=1}^3 (0, L_i)$. These theorems show the independence of the
dynamic  transition  on the nonlinearity of the Onsager mobility.

We consider  the following three cases of the domain:
\begin{subequations}
\begin{align}
L=L_1>L_2>L_3, \label{case1} \\
L=L_1=L_2>L_3, \label{case2} \\
L=L_1=L_2=L_3. \label{case3}
\end{align}
\end{subequations}
The critical temperature at which the homogenous state loses its stability is given by:
\begin{equation} \label{Tc}
T_c=\frac{\overline{u}_1(1-\overline{u}_1)}{R}\left(2\gamma-\frac{\alpha\pi^2}{L^2}\right),
\end{equation}
see Step 2 in Section \ref{sec proof of main thms} for more details. The following numbers, evaluated at $T_c$, are crucial to describe the phase transition of the problem:
\begin{subequations}
\begin{align}
& B_1=\Bigl (b_3-\frac{2L^2}{9\alpha\pi^2}b_2^2\Bigr)\big \vert_{T=T_c}, \label{sigma1}\\
& B_2=\Bigl( b_3-\frac{26L^2}{27\alpha\pi^2}b_2^2\Bigr)\big \vert_{T=T_c},\label{sigma2}\\
& B_3=\Bigl( b_3-\frac{10L^2}{9\alpha\pi^2}b_2^2\Bigr )\big \vert_{T=T_c}. \label{sigma3}
\end{align}
\end{subequations}

\bt \label{thm1}
Assume $L=L_1>L_2>L_3$.
Then the system \eqref{model}--\eqref{meanzero} has a phase transition at $(u,T)=(0,T_c)$. Moreover,
the following statements are true.
\begin{itemize}
\item[i)] If $B_1>0$, then the transition is Type-I. In particular,
the problem bifurcates on $T<T_c$ to exactly two equilibria $u_1^T$
and $u_2^T$ which are attractors and can be expressed as
$$u_{1,2}^T=\pm\sqrt\frac{4R(T_c-T)}{3B_1 \overline{u}_1(1-\overline{u}_1)}\cos \frac{\pi x_1}{L}+o(|T-T_c|).$$
\item[ii)] If $B_1<0$, then the transition is Type-II. In particular,
the problem bifurcates on $T>T_c$ to exactly two equilibria $u_1^T$
and $u_2^T$, which are non-degenerate saddle points given by:
$$u_{1,2}^T=\pm\sqrt {- \frac{4R(T-T_c)}{3B_1 \overline{u}_1(1-\overline{u}_1)} } \cos \frac{\pi x_1}{L}+o(|T-T_c|).$$
\end{itemize}
\et

\bt \label{thm2}
Assume $L=L_1=L_2>L_3$. Then the system \eqref{model}--\eqref{meanzero} undergoes a phase transition at $T=T_c$ satisfying the following properties:
\begin{itemize}
\item[i)] If $B_2>0$,
then the transition is Type-I and the problem bifurcates on $T<T_c$ side to an attractor $\Sigma_T$, which is homeomorphic to the unit sphere $S^1$ and contains 8 non-degenerate singular points with 4 minimal attractors.
\item[ii)] If $B_2<0$,
then the transition is Type-II and the problem bifurcates to 8 non-degenerate saddle points at $T=T_c$. There are 4 saddle points bifurcating out on both sides of $T_c$ if $B_1>0$, and all of the 8 bifurcated saddle points  are on $T>T_c$ side if $B_1<0$.
\end{itemize}
\et

\bt\label{thm3}
Assume $L=L_1=L_2=L_3$. There is a phase transition at $(u, T) = (0, T_c)$ for the system \eqref{model}--\eqref{meanzero}, and the following assertions hold true:
\begin{itemize}
\item[i)] If $B_3>0$, then the phase transition is Type-I, and the problem bifurcates on $T<T_c$ side to an attractor $\Sigma_T$, which is homeomorphic to the unit sphere $S^2$. $\Sigma_T$ contains 26 non-degenerate singular points, among which
\begin{align*}
& \text{8 are minimal attractors \qquad} && \text{if }  b_3<\frac{22L^2b_2^2}{9\pi^2} \text{ at } T=T_c, \text{ and } \\
& \text{6 are minimal attractors \qquad} &&\text{if } b_3> \frac{22L^2b_2^2}{9\pi^2}\text{ at } T=T_c.
\end{align*}
\item[ii)] If $B_3<0$, then the phase transition at $T=T_c$ is Type-II. In particular, the  problem bifurcates to 26 saddles at $T=T_c$. On $T>T_c$, there are
\begin{align*}
& \text{8 saddle points \qquad} && \text{if $B_2>0$ \text{ and } $B_3<0$,} \\
&\text{20 saddle points \qquad} && \text{if $B_1>0$ and $B_2<0$,}\\
&\text{26 saddle points \qquad} && \text{if $B_1<0$,}
\end{align*}
and the rest are on the side when $T<T_c$.
In all these three cases, the saddle points are all non-degenerate.
\end{itemize}
\et

\begin{rem}
When the transition is Type-II, the system undergoes a drastic change as $T$ decreasingly crosses $T_c$. On $T>T_c$, the physically meaningful states are the homogenous state $u=0$ and some transition states away from $u=0$ which are metastable. The bifurcated saddles indicated in the theorems in this case are not physical states.
\end{rem}

\section{Dynamic Transition Framework for Quasilinear Systems} \label{app. CM for quasilinear}
In this section, we present a general framework for studying phase
transitions for quasilinear systems based on the dynamic transition
theory developed recently by Ma and Wang \cite{MW05,MW}. The basic
philosophy is still to search for the complete set of transition states
as in the dynamic transition theory. For quasilinear systems, the
key technical ingredient is the reduction of the original system to
a properly defined center manifold for quasilinear parabolic
equations \cite{Mie91, Simonett95}.

\subsection{Center manifolds for quasilinear systems} \label{CM existence}
Let $X_1 \subset X$ be two Banach spaces with dense and continuous
inclusion. Consider
\begin{equation} \label{Abstract Quasi Eqn.}
\begin{aligned}
&\frac{\mathrm{d}u}{\mathrm{d}t}-L_\lambda u =G(u,\lambda), \\
&u(0)=u_0,
\end{aligned}
\end{equation}
where $u$ is the unknown function in $C([0,\,T];\,X)$, $\lambda$ is
a real parameter of the system, for each $\lambda$ the linear operator $L_\lambda:\,
D(L_\lambda)=X_1 \rightarrow X$ is the infinitesimal generator of an
analytic semigroup $(e^{L_\lambda t})_{t\ge 0}$ with domain $D(L_\lambda)$ independent of $\lambda$, $L_\lambda$ depends continuously on $\lambda$, and $G:\, X_1 \times \mathbb{R}
\rightarrow X$ is a given nonlinear function, which contains terms
of highest order derivatives in space variables and thus makes the problem quasilinear in nature.

As is well known, the starting point of the existence of center
manifolds is the variation of constants formula
\begin{equation} \label{var of const}
u(t)=e^{L_\lambda t}u_0 + \int_0^t e^{L_\lambda (t-s)}G(u(s), \lambda)\,\mathrm{d}s.
\end{equation}

However, this is only a formal expression. To make sense of
\eqref{var of const}, we face two difficulties. First, we need the integral term to be
finite and second, it should be in the same space as $u$.

There is an easy remedy for the first one by strengthening the usual
concept of a solution by requiring
\begin{equation}
u\in C([0,\,T];\, X_1) \cap C^1([0,\,T];\,X).
\end{equation}
This requires, of course, that we choose the initial data $u_0$ in
$X_1$.

To overcome the second difficulty, we have to deal with
the regularity loss due to the nonlinear term $G$. This has to be
compensated by the regularizing properties of the analytic semigroup
generated by the linear part. In order to achieve this, we have to
choose our spaces carefully. As is well known (see e.g. Henry
\cite{Hen81}), for the semilinear case, this can be overcome by
requiring that $G:\,X_\alpha \times \mathbb{R} \rightarrow X$ with
$X_\alpha$ being some intermediate space between $X_1$ and $X$. But
this does not work for the quasilinear case because of the terms
with highest order derivatives involved in $G$. One way to fix this
is to work in a pair of Banach spaces $D_{L_\lambda}(\theta+1)$ and
$D_{L_\lambda}(\theta)$ for some $\theta \in (0,\,1)$ instead of  $X_1$ and $X$,
where $D_{L_\lambda}(\theta+1)$ and
$D_{L_\lambda}(\theta)$ are defined as follows:

\bd \label{Definition D_A}

Let $A$ be the infinitesimal generator of an analytic semigroup
in $X$. For $\theta \in (0,\,1)$, the spaces $D_A(\theta)$ and
$D_A(\theta+1)$ are defined as:
\begin{equation} \label{Def D_A}
\begin{aligned}
&D_A(\theta)=\{\, u\in X\:|\: \|t^{1-\theta} A e^{tA} u\|_X \in L^\infty (0,1), \lim_{t \rightarrow 0^+}\|t^{1-\theta} A e^{tA} u\|_X = 0 \}, \\
&\|u\|_{D_A(\theta)} =\|u\|_X + \max_{0<t<1} \|t^{1-\theta} A e^{tA} u\|_X , \\
&D_A(\theta+1)=\{\, u\in D(A) \:|\: Au \in D_A(\theta) \},\\
&\|u\|_{D_A(\theta+1)} =\|u\|_X + \|Au\|_{D_A(\theta)}.
\end{aligned}
\end{equation}
\ed

The function spaces $D_A(\theta)$ and $D_A(\theta+1)$ are Banach spaces  endowed
with corresponding norms respectively. For any $\theta \in (0,\,1)$,
\begin{equation*}
D_A(\theta)=(X,\, D(A))_{\theta},
\end{equation*}
where $D(A)$ is the domain of $A$, and $(X, Y)_\theta$ is the real
interpolation space between $Y$ and $X$; see e.g. \cite{dPG79,
Lun95, Triebel78}.

It is known that $D_A(\theta)$ does not depend explicitly on the operator $A$, but only on the domain of $A$ and on the graph norm of $A$; see e.g. Corollary 2.2.3 in \cite{Lun95}. So by our assumptions on $L_\lambda$, $D_{L_\lambda}(\theta)$ does not depend on
$\lambda$ as long as $\lambda$ is restricted to some bounded interval in $\mathbb{R}$.
 We refer readers to \cite{Lun95} for some equivalent
characterizations of these two spaces for arbitrary Banach space
$X$. When $X$ is $L^p(\Omega)$ for some properly chosen $p$, these
spaces are contained in the so called (little) Nikolski spaces
$h_p^s(\Omega)$ for some $s$. It is this characterization and the
known nice properties of the Nikolski spaces that help us overcome
the aforementioned second difficulty.

Now, we present the center manifold theorem for \eqref{Abstract
Quasi Eqn.} under the following assumptions:
\begin{itemize} \label{A1A2}
\item[$(A_1)$:] The Banach space $X$ splits into closed
$L_\lambda$-invariant subspaces $E_1^\lambda$ and $E_2^\lambda$ such
that \eqref{Abstract Quasi Eqn.} takes the form
\begin{equation*}
\begin{aligned}
\frac{\mathrm{d}u_c}{\mathrm{d}t} - L_1^\lambda u_c&=P_1G(u_c, u_s, \lambda),\\
\frac{\mathrm{d}u_s}{\mathrm{d}t} - L_2^\lambda u_s&=P_2G(u_c, u_s, \lambda),
\end{aligned}
\end{equation*}
where $u=u_c+u_s,\, u_c\in E_1^\lambda, \,u_s\in X_1\cap E_2^\lambda,\, L_i^\lambda:=L_\lambda|_{E_i^\lambda}$ are the restrictions of $L_\lambda$ to the corresponding invariant subspaces, and $P_i: X \rightarrow E_i^\lambda$ are the canonical projections for $i = 1, \, 2$. Moreover, $\dim E_1^\lambda<\infty$, all eigenvalues of $L_1^\lambda$ have
nonnegative real parts at some $\lambda=\lambda_c$, and for $\lambda$ sufficiently close to $\lambda_c$ the operator
$L_2^\lambda: X_1 \cap E_2^\lambda \rightarrow E_2^\lambda$ is
closed, densely defined and satisfies the resolvent estimate:
\begin{equation*}
\|(L_2^\lambda-z)^{-1}\|_{E_2^\lambda \rightarrow E_2^\lambda} \le
\fraction{C}{1+|z|}, \, \forall \, z \in \mathbb{C} \text{ with } \text{Re}\, z \ge 0.
\end{equation*}

\item[$(A_2)$:] There exist neighborhoods $U_1\subset E_1^\lambda$ and
$U_2 \subset D_{L_2^\lambda}(\theta+1)$ of zero and an integer $k\ge1$
such that
\begin{equation*}
G=(P_1G,\, P_2G) \in C^k_{b, \text{unif}}(U_1\times U_2
\times\mathbb{R}, E_1^\lambda \times D_{L_2^\lambda}(\theta)),
\end{equation*}
where $C^k_{b, \text{unif}}$ is the set of all functions with bounded uniformly continuous derivatives up to order $k$. Moreover, there is a neighborhood $\Lambda$ of $\lambda_c$, such
that $G(0,\, \lambda)=0$, and $(\partial/\partial u)G(0,\,\lambda)=0$
for all $\lambda \in \Lambda$.
\end{itemize}

\bt [\cite{Mie91}]\label{CM thm for Quasilinear}

Let $(A_1)$ and $(A_2)$ be satisfied for \eqref{Abstract Quasi Eqn.}. Then there exist neighborhoods $U_1'
\subset U_1$ and $U_2' \subset U_2$ of zero, a neighborhood
$\Lambda' \subset \mathbb{R}$ of $\lambda_c$, and a function
\begin{equation*}
\Phi=\Phi(u_c,\,\lambda)\in C^k_b(U_1' \times \Lambda', \, U_2')
\end{equation*}
with the following properties:
\begin{itemize}
\item[i)] The set
\begin{equation*}
M_\lambda=\{\, (u_c,\, \Phi(u_c, \,\lambda)) \in E_1^\lambda \times
D(L_2^\lambda)\:|\: u_c\in U_1'\},
\end{equation*}
called the center manifold for
\eqref{Abstract Quasi Eqn.}, is locally invariant, namely for each $u_0\in M_\lambda$,
\begin{equation*}
u_\lambda(t, \, u_0) \in M_\lambda, \quad \forall \: 0\le t <
t_{u_0}.
\end{equation*}
Here $u_\lambda(t,\, u_0)$ is the solution of
\eqref{Abstract Quasi Eqn.} with initial datum $u_0$ and $t_{u_0}$ is some positive constant depending on $u_0$.
\item[ii)] $\Phi(0,\, \lambda)=0$, $(\partial/\partial u_c)\Phi(0,\,
\lambda)=0$.
\end{itemize}

\et

Now we give the definitions and some crucial properties of Nikolski
spaces following \cite{dPG79}, from which we will see that the
assumption $(A_2)$ above can be verified easily when we choose the
spaces carefully.

\bd[\cite{dPG79}]  \label{Def Nikolski}

Let $\sigma \in (0,\,1)$, $p\in(1,\, \infty)$, and $n \in
\mathbb{N}$. Then
$$h_p^\sigma(\mathbb{R}^n) = \{\, u\in L^p(\mathbb{R}^n)\: : \:\,
|t|^{-\sigma} \|u(\cdot + te_j)-u(\cdot)\|_{L_p} \rightarrow 0
\text{ as } t\rightarrow 0, \forall \, j=1,\,\cdots,\, n\},$$ where
$e_j$ is the unit vector in the $j^{th}$ direction.

For $m \in \mathbb{N}$ and any open set $\Omega \subset \mathbb{R}^n$,
\begin{equation*}
\begin{aligned}
h_p^\sigma(\Omega) &= \{\, u\in L^p(\Omega)\:|\: \exists \,
\widetilde{u} \in h_p^\sigma(\mathbb{R}^n) \text{ such that }
\widetilde{u}|_\Omega = u \}, \\
h_p^{m+\sigma}(\Omega) &= \{\, u\in W_p^m(\Omega)\:|\: D^\beta u \in
h_p^\sigma(\Omega), |\beta|=m \}.
\end{aligned}
\end{equation*}

\ed

\bl [\cite{dPG79, Mie91}]\label{crucial lemma} Let $\Omega$ be an open bounded subset of
$\mathbb{R}^n$ with smooth boundary.
\begin{itemize}
\item[i)] For $s>n/p$, $s\notin \mathbb{N}$, the space $h_p^s(\Omega)$ is continuously
embedded in $C(\overline{\Omega})$ and thus forms an algebra.

\item[ii)] For $s=m+\sigma > n/p$, $m\in \mathbb{N}$, $\sigma \in (0,\,1)$,
and $f\in C^{m+k}(\mathbb{R}^l,\, \mathbb{R})$ with some $k\in \mathbb{N}$, the evaluation
mapping
\begin{equation*}
(u_1(\cdot),\, \cdots, \, u_l(\cdot)) \in (h_p^s(\Omega))^l
\rightarrow f(u_1(\cdot),\, \cdots, \, u_l(\cdot)) \in h_p^s(\Omega)
\end{equation*}
is k times continuously differentiable.
\end{itemize}
\el

We note that the first part of our assumption
$(A_2)$ is a direct consequence of Lemma \ref{crucial lemma} under
the condition that $G$ is smooth enough.

\subsection{Approximation of the center manifold function} \label{section App CMF}
In this subsection, we consider an approximation of the center manifold
function $\Phi(x, \lambda)$ for \eqref{Abstract Quasi Eqn.} obtained by Theorem \ref{CM thm for Quasilinear} following the same line as in \cite{MW}.

We assume that the nonlinear term $G(u,\, \lambda)$ in \eqref{Abstract Quasi Eqn.} has the Taylor
expansion about $u=0$ as follows
\begin{equation} \label{G Taylor}
G(u,\, \lambda)= \sum_{m=k}^r G_m(u,\, \lambda) +
o(\|u\|^r_{D_{L_\lambda}(\theta+1)}), \text{ for some } 2\le k\le r,
\end{equation}
where $u \in D_{L_\lambda}(\theta+1),\: G_m:
\underbrace{D_{L_\lambda}(\theta+1) \times \cdots \times
D_{L_\lambda}(\theta+1)}_{m \text{ times}} \rightarrow
D_{L_\lambda}(\theta)$ is an $m$-multiple linear operator,
and $G_m(u,\,\lambda)=G_m(u,\,\cdots \, , u,\, \lambda)$.

Let $\{\, \beta_i(\lambda) \in \mathbb{C} \:|\: i \in \mathbb{N}\}$
be all eigenvalues of $L_\lambda$ counting multiplicities and $\{e_i(\lambda)\:|\: i \in \mathbb{N}\}$ be the corresponding eigenvectors. Assume that the following principle of exchange of stabilities (PES) condition holds:
\begin{equation} \label{Abstract PES}
\begin{aligned}
& \mathrm{Re}\, \beta_i(\lambda)
\begin{cases} <0 & \mbox{if } \lambda < \lambda_c \\ =0 & \mbox{if } \lambda = \lambda_c \\ >0 & \mbox{if } \lambda > \lambda_c
\end{cases} && \forall \, 1\le i\le m,\\
& \mathrm{Re}\, \beta_j(\lambda_c) <0 && \forall \, j\ge m+1,
\end{aligned}
\end{equation}
for some $\lambda_c \in \mathbb{R}$.

We also assume that the span of $\{e_i(\lambda) \:\vert \: i \in \mathbb{N}\}$ is dense in $D_{L_\lambda}(\theta+1)$; namely
\begin{equation} \label{eigen complete}
D_{L_\lambda}(\theta+1) = \overline{span\{e_i(\lambda) \mid i \in \mathbb{N}\}}^{D_{L_\lambda}(\theta+1)}.
\end{equation}

Now, let
\begin{equation*}
\begin{aligned}
E_1^\lambda &=\text{span}\{e_1(\lambda),\,\cdots,\, e_m(\lambda)\},\\
E_2^\lambda &= \text{the complement of } E_1^\lambda \text{ in } X.
\end{aligned}
\end{equation*}
Then $L_\lambda$ is invariant on $E_1^\lambda$ and $E_2^\lambda$, i.e., $L_\lambda$ can be decomposed as
\begin{equation} \label{decomp L}
\begin{aligned}
&L_\lambda = L_1^\lambda \oplus L_2^\lambda, \\
&L_1^\lambda: E_1^\lambda \rightarrow E_1^\lambda, \\
&L_2^\lambda: X_1\cap E_2^\lambda
\rightarrow E_2^\lambda,
\end{aligned}
\end{equation}
where $L_1^\lambda$ is the Jordan matrix of $L_\lambda$ associated with
$\beta_j(\lambda)$ $(1\le i\le m)$, and $L_2^\lambda$ has
eigenvalues $\beta_j(\lambda)$ $(j\ge m+1)$.

Now, we present the following theorem which gives a first
order approximation formula of the center manifold function of
\eqref{Abstract Quasi Eqn.} for $\lambda$ close to $\lambda_c$. The approximation formula is
essential to understand the dynamic behavior of the trivial solution $u\equiv 0$ of \eqref{Abstract Quasi Eqn.} for $\lambda$ near $\lambda_c$.

\bt \label{App CMF}

Assume all the above conditions given in this subsection hold. For the nonlinear term $G(u,
\lambda)$, assume in addition that $(A_2)$ in Subsection \ref{CM existence} holds. Then for $\lambda$ sufficiently close to $\lambda_c$ we have the following approximation for
the center manifold function $\Phi(u_c, \, \lambda)$:
\begin{equation} \label{App CMF formula 1}
\Phi(u_c,\, \lambda)= \int_{-\infty}^0 e^{-\tau L_2^\lambda} P_2G_k(e^{\tau L_1^\lambda} u_c,\, \lambda) \, \mathrm{d}
\tau + o(\|u_c\|_{D_{L_\lambda}(\theta+1)}^k),
\end{equation}
where $L_1^\lambda$ and $L_2^\lambda$ are the linear operators
as given in \eqref{decomp L}, $G_k(u,\, \lambda)$ is the lowest order
$k$-multiple linear operator as in \eqref{G Taylor}, and
$u_c=\sum_{i=1}^m y_ie_i \in D_{L_\lambda}(\theta+1)$ is sufficiently small. In particular,
for some special cases we have the following assertions:
\begin{itemize}
\item[i)]  if $L_1^\lambda$ is diagonal near $\lambda=\lambda_c$, then \eqref{App CMF
formula 1} can be approximated as
\begin{equation} \label{diagonal case}
-L_2^\lambda \Phi(u_c,\, \lambda) = P_2 G_k(u_c,\,\lambda) +o(k).
\end{equation}
Henceforth, $o(k)$ stands for
\begin{equation} \label{little o}
o(k):=o(\|u_c\|_{D_{L_\lambda}(\theta+1)}^k)+O(|\mathrm{Re}\, \beta(\lambda)| \|u_c\|_{D_{L_\lambda}(\theta+1)}^k),
\end{equation}
with $\beta(\lambda)$ being the eigenvalue of $L_
\lambda$ with  largest real part.

\item[ii)] Let $m=2$ and $\beta_1(\lambda)=\overline{\beta_2(\lambda)}=\alpha(\lambda)+
i\rho(\lambda)$ with $\rho(\lambda_c)\neq 0$. If $G_k(u,\, \lambda)$
is bilinear, i.e. $k=2$, then the center manifold function
$\Phi(u_c,\, \lambda)$ can be expressed as
\begin{equation} \label{complex case}
\begin{aligned}
\left[(-L_2^\lambda)^2+ 4\rho^2(\lambda)\right]&
(-L_2^\lambda)\Phi(u_c,\, \lambda) \\
=&\left[(-L_2^\lambda)^2+4\rho^2(\lambda)\right]P_2G_2(u_c,\,\lambda)-2\rho^2(\lambda)P_2G_2(u_c,\,\lambda) \\
&+2\rho^2(\lambda)P_2G_2(y_1e_2-y_2e_1,\,\lambda) \\
&+\rho(\lambda)(-L_2^\lambda)[P_2G_2(y_1e_1+y_2e_2,\,y_2e_1-y_1e_2,\,\lambda) \\
&+P_2G_2(y_2e_1 - y_1e_2, \,y_1e_1+y_2e_2,\,\lambda)] + o(2).
\end{aligned}
\end{equation}

\item[iii)] Let $\beta(\lambda)=\beta_1(\lambda)=\cdots=\beta_m(\lambda)$ have
algebraic multiplicity $m\ge 2$ and geometric multiplicity $r=1$
near $\lambda=\lambda_c$, i.e., $L_1^\lambda$ has the Jordan form:
\begin{equation} \label{multiplicity one}
L_1^\lambda=\left(
    \begin{array}{ccccc}
    \beta(\lambda) & \delta & \cdots &0 & 0\\
    0& \beta(\lambda) & \cdots & 0 & 0 \\
    \vdots& \vdots& \vdots& \vdots& \vdots \\
    0 & 0& \cdots & \beta(\lambda) & \delta \\
    0 & 0& \cdots & 0 & \beta(\lambda)
    \end{array}
    \right) \text{ for some } \delta \neq 0.
\end{equation}

Let
\begin{equation*}
z=\sum_{j=1}^m \xi_je_j \in E_1^\lambda \text{  with  }
\xi_j=\sum_{r=0}^{m-j} \fraction{\delta ^r t^r y_{j+r}}{r!},
\end{equation*}
where $y=(y_1,\, \cdots,\, y_m)\in \mathbb{R}^m$, $\delta$
is as in \eqref{multiplicity one}, and $t\ge 0$. Then there exist functions $F_0(y),\, \cdots, \, F_{k(m-1)}(y)$ such that the $k$-linear term
$G_k(z,\, \lambda)$ can be expressed as
\begin{equation*}
G_k(z,\,\lambda)=F_0(y)+tF_1(y)+\cdots+t^{k(m-1)}F_{k(m-1)}(y),
\end{equation*}
and the center manifold function $\Phi$ has the following form
\begin{equation}
\begin{aligned}
&\Phi=\sum_{j=0}^{k(m-1)} \Phi_j +o(k), \\
&-(L_2^\lambda)^{j+1} \Phi_j= j! P_2F_j(y), \text{
for } 0\le j\le k(m-1).
\end{aligned}
\end{equation}
\end{itemize}
\et

The above Theorem is a direct generalization of the Hilbertian
version in \cite{MW} and the proof is the same as the Hilbertian
version with obvious modification and is thus omitted here.

\section{Proof of Main Theorems on Phase Transitions} \label{sec proof of main thms}
In this section, we provide a unified proof for Theorem \ref{thm1}--\ref{thm3} on the phase transitions of the problem \eqref{model}--\eqref{meanzero}. The main ingredient of our proof is the center manifold reduction,
following the line of Ma and Wang \cite{MW09a}. But since our
equation is quasilinear, it seems very hard,  if not impossible,  to do
the reduction in Hilbert space setting as was done for semilinear case in \cite{MW09a}; see also the discussion in Section \ref{app. CM for quasilinear}. Instead,
we will work with a pair of Banach spaces $(D_{L_T}(\theta+1), D_{L_T}(\theta))$ for some $\theta \in (0,1)$ as defined in Definition \ref{Definition D_A}, where the existence of a center manifold is known and is recalled in Theorem \ref{CM thm for Quasilinear}.

In order to study the phase transition of the problem we need that the equation admits a global solution $u\in C_b([0, \infty); D_{L_T}(\theta+1)) \cap C_b^1([0, \infty); D_{L_T}(\theta))$ at least for small initial data in $D_{L_T}(\theta+1)$. This is done in Section \ref{section well-posedness}, where the existence of global solutions with small initial data in $H^2$ is also shown.

Assuming for the moment the well-posedness of the problem \eqref{model}--\eqref{meanzero} with small initial data in $D_{L_\lambda}(\theta+1)$, we prove the main theorems in five steps. In the first step, we establish the necessary functional set-up. In Step 2, we analyze the linearized problem to identify the critical parameter at which  the homogeneous state $u\equiv 0$ of the system loses its stability. Step 3 is devoted to deriving an approximation of the center manifold function by the approximation formula given in Section \ref{section App CMF}. We derive the reduced equations to center manifolds in Step 4. In the last step, the reduced equation to the corresponding center manifold is  analyzed.

\medskip

{\sc Step} 1: {\it Functional setting.}
For the functional setting of the problem, we will choose $p>3$ and
$\theta>0$ such that $1>4\theta>3/p$ and set
\begin{equation} \label{working spaces}
\begin{aligned}
& D(L_T)=\{u\in W^{4,p}(\Omega)\mid \frac{\partial u}{\partial \nu}=\frac{\partial \Delta u}{\partial \nu}=0,\int_\Omega u\, \mathrm{d}x=0\},\\
& X=\{u\in L^p(\Omega)\mid \int_\Omega u\, \mathrm{d}x=0\}.
\end{aligned}
\end{equation}

With this choice of $p$ and $\theta$, the interpolation
space $D_{L_T}(\theta)$ in \eqref{working interpolation spaces} becomes
 an algebra (see Lemma \ref{crucial lemma}), which is essential
to guarantee the existence of a center manifold. We note that the
algebra property is also needed for the well-posedness; see the proof of Theorem
\ref{existence in interpolation spaces}.

We define the operators $L_T=-A+B_T:D(L_T)\rightarrow X$ by
\begin{equation}
\begin{aligned}
& Au=\alpha H(\overline{u}_1) \Delta^2u, \\
& B_Tu=b_1H(\overline{u}_1)\Delta u,
\end{aligned}
\label{L_T}
\end{equation}
and $G$ by
\begin{equation}
\begin{split}
G(u,T) =& H(\overline{u}_1) \Delta(b_2 u^2+b_3 u^3+o(u^3)) \\
&-H'(\overline{u}_1) \nabla \left [u \nabla(\alpha \Delta u-b_1 u-b_2 u^2+o(u^2))\right ] \\
& -\frac{1}{2}H''(\overline{u}_1) \nabla \left [u^2 \nabla(\alpha
\Delta u-b_1 u+o(u)) \right ].
\end{split}
\label{G}
\end{equation}

The problem \eqref{model}-\eqref{meanzero} can now be recast in the
following abstract form:
\begin{equation}\label{main}
\frac{du}{dt}=L_Tu+G(u,T), \qquad u(0)=\varphi.
\end{equation}

Letting $s=4\theta$, it is known (see \cite{dPG79}) that the interpolation spaces $D_{L_T}(\theta)$ and $D_{L_T}(\theta+1)$ defined in Definition \ref{Definition D_A} are given by
\begin{equation} \label{working interpolation spaces}
\begin{aligned}
& D_{L_T}(\theta)=(X, D(L_T))_\theta =\{u\in h_p^s \mid \int_\Omega u \, \mathrm{d}x =0\}, \\
& D_{L_T}(\theta+1)=\{u\in h^{s+4}_p(\Omega)\mid \frac{\partial
u}{\partial \nu}=\frac{\partial \Delta u}{\partial
\nu}=0,\int_\Omega u\, \mathrm{d}x=0\},
\end{aligned}
\end{equation}
where $h_p^s$ is the Nikolski space defined in Definition \ref{Def
Nikolski}.

From Lemma \ref{crucial lemma}, we know that for $s=m+\sigma>n/p$,
$0<\sigma<1$, $f\in C^{m+k}(\mathbb{R}^l,\, \mathbb{R})$, the
evaluation mapping
\begin{equation*}
(u_1(\cdot),\, \cdots, \, u_l(\cdot)) \in (h_p^s(\Omega))^l
\rightarrow f(u_1(\cdot),\, \cdots, \, u_l(\cdot)) \in h_p^s(\Omega)
\end{equation*}
is k-times continuously differentiable. This immediately implies
that
\begin{equation} \label{smoothness of G}
G(\cdot,T):D_{L_T}(\theta+1)\rightarrow D_{L_T}(\theta) \text{ is smooth for all $T > 0$.}
\end{equation}
We can also check easily that
\be \label{G higher order}
G(0,\, T)=0 \text{ and } \frac{\partial}{\partial u}G(0,\, T)=0 \text{ for all $T>0$.}
\ee

\medskip

{\sc Step} 2: {\it The principle of exchange of stabilities (PES).} In this step, we explore the eigenvalue problem associated with the linearized counterpart of \eqref{main} to identify the critical parameter $T=T_c$ at which the homogeneous state $u\equiv 0$ of the system loses its stability.  First, we consider the eigenvalue problem
\begin{equation}
\begin{aligned}
& -\Delta e_K=\rho_K e_K \qquad \text{in } \Omega, \\
& \frac{\partial e_K}{\partial \nu}|_{\partial\Omega}=0, \\
& \int_\Omega e_K \, \mathrm{d}x=0.
\end{aligned}
\label{eigenvaluelaplacian}
\end{equation}
The eigenvectors $e_K$ and the eigenvalues $\rho_K$ are given by
\begin{equation}
e_K=\prod_{i=1}^3\cos \frac{k_i\pi x_i}{L_i}, \qquad
\rho_K=\sum_{i=1}^3\frac{k_i^2\pi^2}{L_i^2}, \label{rho_K}
\end{equation}
where
\bes 
K\in\mathcal{K}:=\{(k_1,k_2,k_3):k_i\geq 0,\, k_1^2+k_2^2+k_3^2\neq 0\}.
\ees

Now, we turn to the eigenvalue problem associated with the linearization of \eqref{main} around $u\equiv 0$:
\begin{equation}
 L_T e_K=\beta_K(T)e_K.  \label{eigenvalueprob}
 \end{equation}
It is easy to see that the eigenvectors of \eqref{eigenvalueprob} are the
same as the eigenvectors of  \eqref{eigenvaluelaplacian}, and the eigenvalues  are given by
\begin{equation}
\begin{split}
\beta_K(T)& =-H(\overline{u}_1)(\alpha \rho_K^2+\rho_K b_1)\\
&=H(\overline{u}_1)\rho_K\left(2\gamma-\frac{RT}{\overline{u}_1(1-\overline{u}_1)}-\alpha\rho_K\right).
\end{split}
\label{beta_K}
\end{equation}

Let $T_c$ be given by \eqref{Tc}. One can readily see that $\beta_K(T) < 0$ for all $K\in \mathcal{K}$ when $T< T_c$.
Now, we define $\mathcal{P}$, a subset of $\mathcal{K}$, which contains all $K\in \mathcal{K}$ satisfying $\beta_K(T_c) = 0$; namely
\bea \label{P}
\mathcal{P}=
\begin{cases}
\{(1,0,0)\} &\text{ if $L_1>L_2>L_3$}, \\
\{(1,0,0), \, (0,1,0)\} &\text{ if $L_1=L_2>L_3$}, \\
\{(1,0,0), \, (0,1,0), \, (0,0,1)\} &\text{ if $L_1=L_2=L_3$}.
\end{cases}
\eea
By \eqref{rho_K}, \eqref{beta_K} and our choices of $T_c$ and $\mathcal{P}$, we see that PES is valid:
\begin{equation}
\begin{aligned}
& \beta_K(T)
\begin{cases} <0 & \mbox{if } T>T_c, \\ =0 & \mbox{if } T=T_c, \\ >0 & \mbox{if } T<T_c,
\end{cases}
&\qquad  \forall \: K \in \mathcal{P}, \\
& \beta_K (T_c) <0,  &\forall \: K\in  \mathcal{K} \setminus \mathcal{P}.
\end{aligned}
\label{PES}
\end{equation}
The PES above shows that $T_c$ is the critical parameter value at which the homogeneous state loses its linear stability. From the general dynamic transition in \cite{MW}, we know then  that the system will always undergo a dynamic transition at this critical threshold. The type of transitions is however dictated by the nonlinear interactions, which we shall explore in the next few steps.

{\sc Step} 3: {\it Approximation of the center manifold function.}
Let $E_1^T = \text{span}\{e_K \mid K \in \mathcal{P}\}$ and $E_2^T$ be the complement of $E_1^T$ in $X$, where $X$ is defined in \eqref{working spaces}. Let $L_1^T$ and $L_2^T$ be the restrictions of $L_T$ to $E_1^T$ and $E_2^T$, respectively. It is clear that assumption $(A_1)$ below Definition~\ref{Definition D_A} is satisfied for \eqref{main} with $T$ playing the role of $\lambda$. Thanks to \eqref{smoothness of G} and \eqref{G higher order}, $(A_2)$ is also satisfied.
 Thus, by Theorem \ref{CM thm for Quasilinear} the system
\eqref{main} admits a center manifold in a neighborhood of $u=0$ in $D_{L_T}(\theta + 1)$. In the following, we will use Theorem \ref{App CMF} to derive an approximation of the center manifold function $\Phi(u_c, T)$.

Let
\be \label{uc}
u_c=\sum_{J \in \mathcal{P}}y^J e_J,
\ee
where $y^J=y_1^{j_1}y_2^{j_2}y_3^{j_3}$ for $J=(j_1,j_2,j_3)$. Let
\be \label{S}
\mathcal{S}=\{J+L\mid J,L \in \mathcal{P}\}.
\ee
For example if $\mathcal{P}=\{(1,0,0),(0,1,0)\}$ then
$\mathcal{S}=\{(2,0,0), \, (1,1,0), \, (0,2,0) \}$.

Note that the Jordan matrix $L_1^T$ is diagonal for all the three types of domain $\Omega$ as given in \eqref{case1}--\eqref{case3}, then we have the following approximation of the center manifold function $\Phi$ (see Section \ref{section App CMF} formula \eqref{diagonal case}):
\begin{equation} \label{cmformula}
-L_2^T \Phi(u_c, T)= P_2G_2(u_c,T)+o(2),
\end{equation}
where $G_2$ consists of the quadratic terms of $G$ given in \eqref{G}, i.e.,
\begin{equation} \label{G2}
G_2(u_c,T)=H(\overline{u}_1)b_2 \Delta u_c^2-H'(\overline{u}_1)\nabla
(u_c\nabla(\alpha \Delta u_c-b_1 u_c)),
\end{equation}
and the notation $o(n)$ is as in \eqref{little o} with $T$ playing the role of $\lambda$. Henceforth, all the equalities involving $T$ hold for $T$ sufficiently close to $T_c$.

Let $\langle \cdot, \cdot \rangle$ denote the $L_2$ inner product. Note that we have the following orthogonality relations
\be \label{o.g.}
\langle e_J, e_K \rangle \begin{cases} \neq 0, \text{ if } J=K, \\ =0, \text{ if } J\neq K. \end{cases}
\ee
Since $\Phi(u_c, T) \in E_2^T$ and $\{e_K \mid K \in \mathcal{K} \setminus \mathcal{P}\}$ spans $E_2^T$, by the orthogonality relations above, we can write $\Phi$ in the following form:
\be \label{Phi in o.g. basis}
\Phi(u_c, T) = \sum_{K\in \mathcal{K}\setminus \mathcal{P}} \frac{\langle \Phi(u_c, T), e_K \rangle}{\langle e_K, e_K \rangle } e_K.
\ee
Now, for each $e_K$ with $K\in \mathcal{K} \setminus \mathcal{P}$, we take the $L_2$ inner product of \eqref{cmformula} with $e_K$ and integrate by parts on the left hand side to obtain
\begin{equation*} 
\begin{aligned}
\langle -L_2^T \Phi(u_c, T), e_K \rangle &= - \langle \Phi(u_c, T),  L_2^T e_K \rangle \\
&= -\beta_K \langle \Phi(u_c, T),  e_K \rangle=\langle P_2G_2(u_c,T), e_K \rangle +o(2) \\
& = \langle G_2(u_c,T), e_K \rangle +o(2).
\end{aligned}
\end{equation*}
The last equality above holds due to \eqref{o.g.}. Thus,
\bes
\langle \Phi(u_c, T),  e_K \rangle=- \frac{\langle G_2(u_c,T), e_K \rangle}{\beta_K} +o(2).
\ees
Plugging this back to \eqref{Phi in o.g. basis}, we obtain
\begin{equation} \label{Phi in o.g. basis 2}
\begin{aligned}
\Phi(u_c, T) = -\sum_{K\notin \mathcal{P}}  \frac{\langle G_2(u_c,T), e_K \rangle}{\beta_K \langle e_K, e_K \rangle}e_K + o(2).
\end{aligned}
\end{equation}
Also note that for all $K\in \mathcal{P}$, $\beta_K(T) \rightarrow 0$ as $T \rightarrow T_c$.
Then by \eqref{rho_K} and \eqref{beta_K}, we have
\begin{equation} \label{O(beta)}
\alpha \rho_K+b_1=O(\beta_K(T))  \text{ as } T \rightarrow T_c \text{ for all } K\in \mathcal{P}.
\end{equation}

We now compute the term $\langle G_2(u_c,T), e_K \rangle$. By \eqref{uc}, we have
\begin{equation} \label{center.01}
\begin{aligned}
& \langle \Delta u_c^2, e_K \rangle = \langle  u_c^2, \Delta e_K \rangle = -\rho_K \sum_{J,L\in \mathcal{P}} \int_\Omega y^{J+L}  e_J e_L e_K \d x,
\end{aligned}
\end{equation}
and
\begin{equation} \label{center.02}
\begin{aligned}
\langle \nabla \cdot (u_c \nabla(\alpha \Delta u_c  & -  b_1 u_c)),e_K \rangle  \\
& = - \langle \sum_{J \in \mathcal{P}}y^Je_J \nabla(\sum_{L\in \mathcal{P}}y^L(  \alpha \Delta e_L -b_1 e_L)) , \nabla e_K \rangle  \\
& = \langle \sum_{J \in \mathcal{P}}y^Je_J \nabla(\sum_{L\in \mathcal{P}}y^L(  \alpha \rho_L  + b_1) e_L) , \nabla e_K \rangle \\
& = \sum_{J,L \in \mathcal{P}}(  \alpha \rho_L  + b_1)y^{J+L} \int_\Omega e_J  \nabla e_L \nabla e_K \d x.
\end{aligned}
\end{equation}
By our definitions of $\mathcal{P}$ and $\mathcal{S}$ in \eqref{P} and \eqref{S}, respectively, one can easily see that for any given $J,\, L \in \mathcal{P}$ and $K \in \mathcal{K} \setminus \mathcal{P}$, we have:
\bea \label{center.03}
& \int_\Omega  e_J e_L e_K \d x = \begin{cases} \frac{1}{4}V & \text{ if } K=J+L, \\
0 & \text{ otherwise,}
\end{cases} \\
& \int_\Omega e_J  \nabla e_L \nabla e_K \d x  \begin{cases} \neq 0 & \text{ if } K=J+L, \\
=0&  \text{ otherwise.}
\end{cases}
\eea
Here $V=L_1L_2L_3$ is the volume of $\Omega$.

Now, by \eqref{G2}, and \eqref{center.01}--\eqref{center.03}, we have
\be \label{center.04}
\langle G_2(u_c,T),e_K \rangle = 0, \Forall K \in \mathcal{K} \setminus (\mathcal{P} \cup \mathcal{S}).
\ee

By \eqref{O(beta)} and \eqref{center.02}, we also have
\begin{equation} \label{center.05}
\begin{aligned}
& \langle \nabla \cdot (u_c \nabla(\alpha \Delta u_c-b_1 u_c)),e_K \rangle = o(2), \Forall K \in \mathcal{S}.
\end{aligned}
\end{equation}
Hence by \eqref{Phi in o.g. basis 2}, \eqref{center.01}, \eqref{center.04} and \eqref{center.05}, the center manifold has the following approximation:
\begin{equation}\label{center.1}
\begin{aligned}
& \Phi(y)=\sum_{K\in \mathcal{S}} \Phi_K e_K +o(2), \\
& \Phi_K= \frac{\langle H(\overline{u}_1)b_2 \Delta u_c^2, e_K\rangle}{-\beta_K \langle
e_K, e_K\rangle} =\frac{ H(\overline{u}_1)b_2\rho_K}{\beta_K
\langle e_K, e_K\rangle} \sum_{J,L\in \mathcal{P}}y^{J+L}
\int_\Omega e_J e_L e_K \d x, \: K  \in \mathcal{S}.
\end{aligned}
\end{equation}
Using \eqref{beta_K}, \eqref{O(beta)} and \eqref{center.03}, we have
\begin{equation} \label{center.2}
\begin{aligned}
& \Phi_{2J}=- \frac{b_2y^{2J}}{6\alpha \rho_J} +O(\beta_1(T)|y^{2J}|), && J\in \mathcal{P},  \\
& \Phi_{J+L}= - \frac{2 b_2y^{J+L}}{\alpha \rho_J} +O(\beta_1(T)|y^{J+L}|), && J\neq L
\text{ and } J,\: L \in \mathcal{P}.
\end{aligned}
\end{equation}

{\sc Step} 4: {\it Derivation of the reduced system.}
Now let
\be \label{decomp u}
u=\sum_{J \in \mathcal{P}} y^Je_J + \Phi(y, T).
\ee
The dynamics
of the system \eqref{main} close to  $T_c$ is determined by the dynamics on the corresponding center manifold. To this end, we replace $u$ in \eqref{main} by the right hand side of \eqref{decomp u}, take the $L_2$ inner product of \eqref{main} with $e_J$, and make use of the orthogonality relations \eqref{o.g.} to obtain the following reduced system:
\begin{equation} \label{reduced.1}
\frac{dy^J}{dt}=\beta_J(T)y^J+ \frac{\langle
G(u,T),e_J\rangle}{\langle e_J,e_J \rangle}, \qquad J \in
\mathcal{P}.
\end{equation}
The second term on the right hand side of \eqref{reduced.1} can be simplified further using the approximation formula of the center manifold function \eqref{center.2} as we now show. For $J\in \mathcal{P}$, making use of the
orthogonality relations \eqref{o.g.}, the following can be obtained by direct computation:
\begin{equation} \label{comp.1}
\begin{split}
 \langle \Delta u^2,e_J \rangle &=-\rho_J \langle u^2,e_J\rangle=-2\rho_J \sum_{L \in \mathcal{P}, K \in \mathcal{S}} y^{L}\Phi_K \int_\Omega e_L e_K e_J\d x+o(3) \\
 &=- \frac{V\rho_J }{2} \sum_{L\in \mathcal{P}} y^{L}
 \Phi_{J+L}+o(3).
 \end{split}
\end{equation}
\begin{equation}\label{comp.2}
\begin{split}
 \langle \Delta u^3,e_J \rangle &=-\rho_J \sum_{K,L,M\in \mathcal{P}} y^{K+L+M} \int_\Omega e_Ke_Le_Me_J \d x+o(3) \\
 & =-\rho_J (y^{3J} \int_\Omega e_J^4 \d x + 3 \sum_{L \in \mathcal{P}, L \neq J} y^{J+2L} \int_\Omega e_J^2 e_L^2 \d x ) +o(3) \\
 & =- \frac{3V\rho_J }{8}\Bigl(y^{3J}  + 2 \sum_{L \in \mathcal{P}, L \neq J} y^{J+2L} \Bigr )+o(3).
 \end{split}
\end{equation}
\begin{equation} \label{comp.3}
\begin{split}
 \langle \nabla \cdot(u & \nabla(\alpha \Delta u- b_1 u), e_J \rangle \\
 =& \sum_{L\in \mathcal{P}, K \in \mathcal{S}} y^{L} \Phi_K (\alpha \rho_K+b_1) \int_\Omega e_L \nabla e_K \nabla e_J \d x+o(3) \\
 =&\sum_{L \in \mathcal{P}} y^{L} \Phi_{J+L} (\alpha \rho_{J+L}+b_1) \int_\Omega e_L \nabla e_{J+L} \nabla e_J \d x+o(3) \\
 =&\frac{V\rho_J}{4} \Bigl[2y^{J} (\alpha \rho_{2J}+b_1)\Phi_{2J}  +\sum_{L \in \mathcal{P},L \neq J} y^{L} \Phi_{J+L} (\alpha \rho_{J+L}+b_1) \Bigr] +o(3) \\
  =&\frac{V}{4}\alpha \rho_J^2 \bigl [6 y^{J} \Phi_{2J}+\sum_{L \in \mathcal{P},L \neq J} y^{L} \Phi_{J+L} \bigr]
  +o(3).
 \end{split}
\end{equation}
\begin{equation} \label{comp.4}
\begin{split}
\langle \nabla \cdot(u \nabla &  u^2),e_J \rangle \\
=& \sum_{K,L,M\in \mathcal{P}} y^{K+L+M} \ \int_\Omega e_Ke_L\nabla e_M \cdot \nabla e_J \d x+\langle u^3, \Delta e_J \rangle +o(3) \\
=& \sum_{L \in \mathcal{P}} y^{J+2L} \int_\Omega e_L^2 |\nabla e_J^2| \d x+\langle u^3,\Delta e_J \rangle +o(3)\\
=& \frac{V\rho_J}{8}(y^{3J} +  2 \sum_{L \in \mathcal{P}, L \neq J} y^{J+2L})+\langle u^3,\Delta e_J \rangle +o(3)\\
=& - \frac{V \rho_J}{4} (y^{3J}+ 2 \sum_{L \in \mathcal{P}, L \neq J}
y^{J+2L})+o(3).
\end{split}
\end{equation}
The last equality above follows from the result of \eqref{comp.2}.
\bea \label{comp.5}
\langle \nabla \cdot ( & u^2\nabla(\alpha\Delta
u -  b_1u)),e_J\rangle \\
&=- \langle \sum_{K,L\in \mathcal{P}} y^{K+L}e_Ke_L\nabla(\sum_{M\in \mathcal{P}}y^M(\alpha\Delta
e_M-b_1e_M)), \nabla e_J\rangle +  o(3) \\
&= \sum_{K,L,M\in \mathcal{P}} y^{K+L+M}(\alpha \rho_M + b_1) \int_\Omega e_Ke_L\nabla e_M \nabla e_J \d x +  o(3) \\
& = o(3) \quad (\text{by } \eqref{O(beta)}).
\eea
Using \eqref{center.1}--\eqref{center.2} and \eqref{comp.1}--\eqref{comp.5} in \eqref{reduced.1} and $\rho_J = \frac{\pi^2}{L^2}$ for all $J \in \mathcal{P}$ with $L$ as in \eqref{case1}--\eqref{case3}, we get the following reduced system:
\begin{equation} \label{reduced.2}
\frac{dy^J}{dt}=\beta_J(T) y^J - \frac{H(\overline{u}_1) \pi^2}{2L^2}
y^J(\sigma_1 y^{2J} +\sigma_2 \sum_{\substack{L \in
\mathcal{P}\\ L\neq J}}y^{2L})+o(3), \quad J\in\mathcal{P},
\end{equation}
where $\sigma_1$ and $\sigma_2$ are
\bea \label{sigma}
\sigma_1 = \frac{3b_3}{2} - \frac{L^2b_2^2}{3\alpha \pi^2}, \text{ and }
\sigma_2 &= 3b_3 - \frac{4L^2b_2^2}{\alpha \pi^2}.
\eea

\medskip

{\sc Step} 5: {\it Analysis of the reduced system.}
The reduced equation \eqref{reduced.2} is essentially the same as in the case of
constant mobility except for a factor
of $H(\overline{u}_1)$ appearing in the cubic terms; see Ma and Wang \cite{MW09a}. For the sake of completeness, we present here the main ingredients of the analysis.

\medskip

{\sc First}, it is known that the transition type of \eqref{reduced.2} at
the critical point $T_c$ given by \eqref{Tc} is completely determined
by the following equations:
\begin{equation} \label{reduced.2critical}
\frac{dy^J}{dt}=-\frac{H(\overline{u}_1)\pi^2}{2L^2}y^J\Bigl( \sigma^0_1y^{2J}+\sigma^0_2\sum_{\substack{L\in \mathcal{P} \\ L\neq J}}y^{2L}\Bigr )\qquad \Forall J \in \mathcal{P},
\end{equation}
where
\bea
\sigma_1^0 = \sigma_1\vert_{T=T_c}, \text{ and }
\sigma_2^0 &= \sigma_2\vert_{T=T_c}.
\eea

Recall $B_1$, $B_2$ and $B_3$ given in \eqref{sigma1}--\eqref{sigma3}. It is easy to see that
\begin{equation} \label{8.92}
\left.
\begin{aligned}
&\sigma^0_1>0\Leftrightarrow B_1 > 0, && \sigma^0_1 <0\Leftrightarrow B_1<0, \\
&\sigma^0_1+\sigma^0_2>0\Leftrightarrow B_2 > 0, && \sigma^0_1+\sigma^0_2<0\Leftrightarrow B_2<0, \\
&\sigma^0_1+2 \sigma^0_2>0\Leftrightarrow B_3 > 0, && \sigma^0_1+2\sigma^0_2<0\Leftrightarrow B_3<0.
\end{aligned}
\right.
\end{equation}
These relations will be used frequently in the following.

\medskip

{\sc Second}, for the case where  $L=L_1>L_2\geq L_3$, the critical index set $\mathcal{P}=\{(1,0,0)\}$, the equation \eqref{reduced.2} reads:
\begin{equation} \label{1inter}
\frac{dy_1}{dt}=\beta_{(1,0,0)}y_1 - \frac{H(\overline{u}_1)\pi^2}{2L^2} \sigma_1^0 y_1^3 + o(3),
\end{equation}
 and \eqref{reduced.2critical} takes the following form:
\begin{equation} \label{1inter critical}
\frac{dy_1}{dt}= - \frac{H(\overline{u}_1)\pi^2}{2L^2}
\sigma_1^0 y_1^3.
\end{equation}
Thus, the system has a pitchfork bifurcation at $T_c$, and the type of transition depends on the sign of $\sigma_1^0$. If $\sigma_1^0 > 0$, namely $B_1>0$, then the bifurcation happens on the side when $T<T_c$, the bifurcated two steady states are local attractors, and the transition is Type-I. If $\sigma_1^0 < 0$, namely $B_1 < 0$, the bifurcation happens on the side when $T>T_c$, the bifurcated two steady states are both saddle points, and the transition is Type-II. It is clear now that the  assertions in Theorem \ref{thm1} hold true.

\medskip

{\sc Third}, for the  case  where  $L=L_1 = L_2 > L_3$, the equations in \eqref{reduced.2}
read:
\begin{equation}\label{2inter}
\begin{aligned}
& \frac{dy_1}{dt}=\beta_{(1,0,0)}(T)y_1- \frac{H(\overline{u}_1) \pi^2}{2L^2}
y_1(\sigma_1^0 y_1^2 +\sigma_2^0 y_2^2) + o(3), \\
& \frac{dy_2}{dt}=\beta_{(0,1,0)}(T)y_2 - \frac{H(\overline{u}_1) \pi^2}{2L^2} y_2(\sigma_1^0
y_2^2 +\sigma_2^0 y_1^2) + o(3),
\end{aligned}
\end{equation}
and the equations in \eqref{reduced.2critical} read:
\begin{equation}\label{2inter critical}
\begin{aligned}
& \frac{dy_1}{dt}=- \frac{H(\overline{u}_1) \pi^2}{2L^2}
y_1(\sigma_1^0 y_1^2 +\sigma_2^0 y_2^2), \\
& \frac{dy_2}{dt}=- \frac{H(\overline{u}_1) \pi^2}{2L^2} y_2(\sigma_1^0
y_2^2 +\sigma_2^0 y_1^2).
\end{aligned}
\end{equation}
To analyze \eqref{2inter critical}, we first find the straight line orbits, which are orbits of the form $y_2 = m_1 y_1$ or $y_1 = m_2y_2$.

We assume that the line
\bes
y_2 = m_1 y_1
\ees
is a straight line orbit of \eqref{2inter critical} with some $m_1\in \mathbb{R}$. Then
\begin{equation}
\begin{aligned}
&\frac{dy_2}{dy_1}=m_1=m_1\frac{\sigma^0_1m^2_1+\sigma^0_2}{\sigma^0_1+\sigma^0_2m^2_1}.
\end{aligned}
\end{equation}
Thus $m_1 = 0,\: \pm 1$ provided $\sigma^0_1 \neq \sigma_2^0$. Similarly, in order that $y_1 = m_2 y_2$ be a straight line orbit of \eqref{2inter critical}, $m_2$ can only take the values $0,\: \pm 1$ provided $\sigma^0_1 \neq \sigma_2^0$.

There are four straight lines in total determined by $y_2 = m_1y_1$ and $y_1 = m_2y_2$ with $m_1,\: m_2 = 0, \: \pm 1$, and each of them contains two orbits. Hence, the system \eqref{2inter critical} has exactly eight straight line orbits provided that $\sigma^0_1 \neq \sigma_2^0$.

Since \eqref{2inter} is a gradient-type equation, the energy decreases along the orbits. Therefore
there are no elliptic regions at $y=0$. Hence, when
$\sigma^0_1+\sigma^0_2>0$ and $\sigma^0_1 \neq \sigma^0_2$ all the straight line orbits tend to $y=0$ which implies that the regions are parabolic and
stable, therefore $y=0$ is asymptotically stable for \eqref{2inter}.
Accordingly, by the attractor bifurcation theorem, Theorem 6.1 in \cite{MW05}, the transition of \eqref{2inter} at $T_c$ is Type-I.

When $\sigma^0_1=\sigma^0_2$, one can check directly that $\sigma^0_1=\sigma^0_2>0$. In this case, it is clear that
$y=0$ is an asymptotically stable singular point of \eqref{2inter critical}.
Hence, the transition of \eqref{2inter} at $T_c$ is
Type-I.

When $\sigma^0_1+\sigma^0_2<0$ and $\sigma^0_1>0$, namely $B_1<0 \text{ and } B_2 > 0$,
the four straight line orbits on $y_2 = \pm y_1$ extend outward from
$y=0$, and the other four on $y_1 = 0$  or $y_2 = 0$ go toward $y=0$ which implies that all regions at $y=0$ are
hyperbolic. Hence, by Theorem $A.3$ in \cite{MW09a},  the transition of \eqref{2inter} at $T_c$ is Type-II.

When $\sigma^0_1\leq 0$, then $\sigma^0_2<0$ too. In this case, no
orbits of \eqref{2inter critical} go toward $y=0$ which implies by Theorem $A.3$ in \cite{MW09a} that the transition is Type-II.

Thus by (\ref{8.92}) and the above analysis, we proved that the transition of \eqref{3inter} from $(u,\, T)=(0,\,T_c)$ is
Type-I if $B_2>0$, and Type-II
if  $B_2<0$. This proves the assertions about the types of transitions stated in Theorem \ref{thm2}.

\medskip

{\sc Fourth}, for the case where  $L=L_1 = L_2 =L_3$,  the equations in \eqref{reduced.2} read
\begin{equation} \label{3inter}
\left.
\begin{aligned}
&\frac{dy_1}{dt}=\beta_{(1,0,0)}(T)y_1-y_1[\sigma^0_1y^2_1+\sigma^0_2(y^2_2+y^2_3)] + o(3),\\
&\frac{dy_2}{dt}=\beta_{(0,1,0)}(T)y_2 -y_2[\sigma^0_1y^2_2+\sigma^0_2(y^2_1+y^2_3)] + o(3),\\
&\frac{dy_3}{dt}=\beta_{(0,0,1)}(T)y_3 -y_3[\sigma^0_1y^2_3+\sigma^0_2(y^2_1+y^2_2)] + o(3),
\end{aligned}
\right.
\end{equation}
and \eqref{reduced.2critical} are
written as
\begin{equation} \label{3inter critical}
\left.
\begin{aligned}
&\frac{dy_1}{dt}=-y_1[\sigma^0_1y^2_1+\sigma^0_2(y^2_2+y^2_3)],\\
&\frac{dy_2}{dt}=-y_2[\sigma^0_1y^2_2+\sigma^0_2(y^2_1+y^2_3)],\\
&\frac{dy_3}{dt}=-y_3[\sigma^0_1y^2_3+\sigma^0_2(y^2_1+y^2_2)].
\end{aligned}
\right.
\end{equation}
It is clear that the straight lines
\bea \label{8.98}
& y_i=0, \ y_j=0\qquad  \hspace{0.6em} \text{ for } i\neq j,\ 1\leq i,\ j\leq
3, \\
& y^2_i=y^2_j,\  y_k=0 \qquad \text{ for } i\neq j, \ i\neq
k,\ j\neq k,\ 1\leq i,j,k\leq 3,\\
& y^2_1=y^2_2=y^2_3,
\eea
consist of orbits of (\ref{3inter critical}). There are 13 straight
lines in total contained in (\ref{8.98}), each of which consists of
two orbits. Thus, (\ref{3inter critical}) has at least 26 straight line
orbits. In fact, as shown in \cite{MW09a}, the number of straight line orbits of (\ref{3inter critical}) is exactly 26 when $\sigma^0_1\neq\sigma^0_2$.

As before, when $\sigma^0_1=\sigma^0_2$, we have that
$\sigma^0_1=\sigma^0_2>0$. In this case, it is clear that
$y=0$ is an asymptotically stable singular point of (\ref{3inter critical}).
Hence, the transition of \eqref{3inter} at $T_c$ is
Type-I.

When $\sigma^0_1+2 \sigma^0_2>0$ and $\sigma^0_1\neq\sigma^0_2$, all
straight line orbits of (\ref{3inter critical}) go toward $y=0$, which
implies  that the regions at $y=0$, are stable, and
$y=0$ is asymptotically stable. Thereby the transition of
\eqref{3inter} is Type-I.

When $\sigma^0_1+2 \sigma^0_2<0$ and $\sigma^0_1 + \sigma^0_2>0$, we can check that $\sigma_1^0 \neq \sigma_2^0$ and hence all straight line orbits of \eqref{3inter critical} are given by \eqref{8.98}. Moreover, all the straight line orbits determined by $y^2_1=y^2_2=y_3^2$ extend outward the origin, and all the rest straight line orbits go toward the origin. Hence, for any initial data in a small neighborhood of $0$, the orbit of \eqref{3inter critical} goes away from $0$ as long as the initial data does not belong to any of the coordinate planes, which implies that the transition is Type-II.

Similarly, when $\sigma^0_1+\sigma^0_2 \le 0$ one can also check that given a small neighborhood of $0$, there is a dense subset of the neighborhood, such that for any initial data in the dense subset, the orbit of \eqref{3inter critical} goes away from $0$. Hence, the transition is Type-II.

Thus by (\ref{8.92}) and the above analysis we proved that the transition of \eqref{3inter} from $(u,\, T)=(0,\,T_c)$ is
Type-I if $B_3>0$, and Type-II
if  $B_3<0$. This proves the assertions about the types of transitions stated in Theorem \ref{thm3}.

\medskip

{\sc Fifth,} we show the nondegeneracy of bifurcated steady states.
Since the bifurcated equilibrium points of \eqref{main} are in one-to-one correspondence to the bifurcated equilibrium points of \eqref{reduced.2}, it is sufficient to consider the leading order steady state equations of the reduced system (\ref{reduced.2})
\begin{equation}
\beta_J(T)y^J- y^J(a_1y^{2J}+a_2\sum_{\substack{L \in \mathcal{P} \\ L\neq J}}y^{2L})=0\qquad
\text{ for }  J\in \mathcal{P},\label{8.101}
\end{equation}
where $a_1=H(\overline{u}_1)\pi^2\sigma_1/(2L^2), a_2=H(\overline{u}_1) \pi^2\sigma_2/(2L^2)$.

Let $m = |\mathcal{P}|$. In \cite{MW09a}, it is shown that (\ref{8.101}) has $3^m-1$ bifurcated solutions, and all bifurcated solutions of (\ref{8.101})
are regular.

For Type-I transition case, since all bifurcated singular points of \eqref{main} are non-degenerate and when $\Sigma_T$ is
restricted to $y_iy_j$-plane $(1\leq i,j\leq m)$ the singular
points are connected by their stable and unstable manifolds, all singular points in $\Sigma_T$ are
connected by their stable and unstable manifolds. Therefore,
$\Sigma_T$ must be homeomorphic to a sphere $S^{m-1}$.

{\sc Finally}, in addition, as in \cite{MW09a}, the number of minimal attractors is obtained by studying the Jacobian matrix of \eqref{8.101}. The proofs of Theorems \ref{thm1}--\ref{thm3} are now complete.

\section{Existence and Uniqueness of Global Strong Solutions} \label{section well-posedness}
In this section, we will give two results concerning the existence
and uniqueness of solutions with small initial data, one in Hilbert
space setting and the other in the interpolation space setting.

\subsection{Existence in Hilbert spaces}
We start with the following problem:
\begin{equation} \label{CHE recall}
\begin{split}
&\frac{\partial u}{\partial t}= \nabla \cdot \left[ H(u) \nabla
(-\alpha \Delta u + b_1 u + b_2 u^2 + b_3 u^3) \right] \\
& \frac{\partial u}{\partial \nu}|_{\partial \Omega}=0, \qquad
\frac{\partial \Delta u}{\partial \nu}|_{\partial \Omega}=0, \\
&u(0)=u_0,\\
&\int_\Omega u\, \mathrm{d}x=0.
\end{split}
\end{equation}
Here $\alpha, b_1, b_2$, and $b_3$ are constants with $\alpha>0$ and $b_3>0$, and $\Omega$ is a bounded domain in $\mathbb{R}^3$ with sufficient smooth boundary.

We make the following
assumption on the Onsager mobility $H(s)$:
\begin{itemize}
\item[($\mathcal{H}$):] $\min H(s)\ge B_1>0$, and $H(s)$ and $H'(s)$ satisfy the following growth condition:
\begin{equation*}
|H(s)| \le C (|s|^{p+1}+1), \: |H'(s)| \le C(|s|^{p}+1) \quad
\forall \: s\in \mathbb{R},
\end{equation*}
\end{itemize}
where $1<p<3$.

It is clear that the free energy functional associated with \eqref{CHE recall} takes the following form (see Section \ref{section the model}):
\begin{equation} \label{new G}
\mathcal{G}(u) =\int_\Omega \frac{\alpha}{2}|\nabla u|^2 + \frac{1}{2}b_1u^2 +
\frac{1}{3}b_2u^3 + \frac{1}{4}b_3u^4\, \mathrm{d}x,
\end{equation}
and the generalized chemical potential $\mu$ in this case is given by
\begin{equation} \label{new mu}
\mu := \frac{\delta \mathcal{G}}{\delta u}= -\alpha \Delta u + b_1u + b_2u^2 + b_3u^3.
\end{equation}

We use the following notations. $|\cdot|$ denotes either the norm on
$L^2(\Omega)$ or the Euclidean norm on $\mathbb{R}^n$, which should
be clear from the context, $|\cdot|_X$ denotes the norm on the
generic Banach space $X$, $\langle \cdot, \, \cdot \rangle$ is
the $L^2(\Omega)$ inner product, $H^m$ is the usual Sobolev space,
and we also denote:
\begin{align}
&W:=\{\, w\in H^2(\Omega)\:|\: \frac{\partial w}{\partial \nu}|_{\partial \Omega} =0 \text{ and }\int_\Omega w\, \mathrm{d} x =0\, \}, \label{W notation} \\
&W_1 := \{\, w\in H^4(\Omega)\:|\: \frac{\partial w}{\partial \nu}|_{\partial \Omega}=\frac{\partial \Delta w}{\partial \nu}|_{\partial \Omega} =0 \text{ and } \int_\Omega w\, \mathrm{d} x =0\, \},  \label{W_1}\\
&\mathcal{A}(t):=  |u(t)|_{H^2}^2+1 \label{A(t) notation}.
\end{align}

Hereafter, $C$ denotes a generic constant which
depends only on the bound $B_1$, the coefficients $b_1,\, b_2,\,
b_3$, and the domain $\Omega$, $C(u_0)$ denotes a generic constant
depending on the initial data $u_0$.

We have the following existence and uniqueness theorem of
a strong solution to the problem \eqref{CHE recall}, which will be proved in Section~\ref{6.3}.

\bt  \label{Global well-posedness}
There exists a constant $\epsilon_0 > 0$, such that for any  initial datum $u_0 \in W$ with $|u_0|_{H^2} < \epsilon_0$, there exists a unique strong solution $u$ to \eqref{CHE recall} such that
$$ u\in L^2(0,\,T; \,W_1) \cap
C([0,\, T];\, W) \text{ with } \frac{\mathrm{d} u}{\mathrm{d} t} \in L^2(0,T; L^2(\Omega)) \quad \forall \: T>0.$$

\et

\subsection{Existence in interpolation spaces}
Now recall the Cahn-Hilliard equation with Onsager mobility:
\begin{equation} \label{CHE. recall}
\begin{aligned}
&\frac{\mathrm{d} u}{\mathrm{d} t}=L_T u + G(u,T), \\
& u(0)=u_0,
\end{aligned}
\end{equation}
where $L_T$ is as in \eqref{L_T} and $G$ is as in \eqref{G}.
The  main result  for \eqref{CHE. recall} is as follows:

\bt \label{existence in interpolation spaces}

Let $D_{L_T}(\theta)$ and $D_{L_T}(\theta + 1)$ be as in
\eqref{working interpolation spaces}, with some $p>3$ and $\theta>0$
such that $1>4\theta>3/p$. Then $\exists \: \epsilon
>0$ and $r > 0$ such that $\forall \: T \ge T_c
- \epsilon$, $\forall \: u_0 \in B(0,r) \subset D_{L_T}(\theta +
1)$, the equation \eqref{CHE. recall} has a unique strong solution
$u \in C_b([0,\,\infty);\, D_{L_T}(\theta+1)) \cap
C_b^1([0,\,\infty);\, D_{L_T}(\theta))$ with $u(0)=u_0$.

\et
The proof of this theorem will be given in Section~\ref{6.4}.
\subsection{Proof of Theorem~\ref{Global well-posedness}}\label{6.3}

The proof is carried out by first proving a local existence result. For this purpose, we need the following lemmas.

\bl \label{equi norm lemma}
$|\Delta u|$ is a norm on $W$ which is equivalent to the $H^2$-norm. Similarly,
$|\Delta ^2 u|$ is a norm on $W_1$ which is equivalent to the $H^4$-norm. Moreover, for any $u\in W_1$, there exists a constant $C$ depending only on the domain $\Omega$ such that
\begin{equation}
|u|_{H^3} \le C |\nabla \Delta u|.
\end{equation}
\el

\bp The above results follow from the regularity theory for elliptic boundary-value problems. For the first claim, we use the regularity theory of the Neumann problem
$$\Delta u = h \mbox{ in } \Omega, \quad \frac{\partial u}{\partial n}\vert_{\partial \Omega} =0,$$
which implies that
$$|u-(u)_{\Omega}|_{H^2} \le C |h| = C|\Delta u|,$$
where $(u)_{\Omega} = \frac{1}{|\Omega|} \int_\Omega u \, \mathrm{d}x$. Since each $u \in W$ satisfies $\int_\Omega u \, \mathrm{d}x = 0$, the first claim follows.

The second claim follows from the regularity theory of the Neumann biharmonic problem
$$\Delta^2 u= h \mbox{ in } \Omega, \quad \frac{\partial u}{\partial n}\vert_{\partial \Omega} =\frac{\partial \Delta u}{\partial n}\vert_{\partial \Omega} =0,$$
which implies that
$$|u-(u)_{\Omega}|_{H^4} \le C |h| = C|\Delta^2 u|.$$
For more details, we refer the interested readers to \cite{Tem97}, Chapter III Lemma 4.2.

For the third claim we use a special case of Corollary 27 in \cite{Dang95}, which states that if $u\in H^3(\Omega)$ and  $\frac{\partial u}{\partial n}\vert_{\partial \Omega} =0$, then there exists a constant $C$ depending only on $\Omega$ such that
$$|u|_{H^3} \le C|\Delta u|_{H^1}.$$
Thus, for any $u\in W_1$,
\begin{equation*}
\begin{aligned}
|u|_{H^3} & \le C|\Delta u|_{H^1}
\le C (|\nabla \Delta u| + |\Delta u|)
\le C |\nabla \Delta u|.
\end{aligned}
\end{equation*}
The last inequality follows by applying the Poincar\'e's inequality to $\Delta u$ and making use of the fact that $\Delta u$ has mean zero due to Gauss divergence theorem and $\frac{\partial u}{\partial n}\vert_{\partial \Omega}=0$.

\ep

\bl  \label{known energy lemma}

Let $u(t)$ be a solution to \eqref{CHE recall} with
initial data $u_0\in W$. Then we have the following estimates:
\begin{equation} \label{G bound}
\mathcal{G}(u(t)) \le C|u_0|_{H^2}^2(1+|u_0|_{L^2}^2), \quad
\forall \: t \ge 0,
\end{equation}
\begin{equation}  \label{H^1 uniform bound}
 |u(t)|_{H^1} \le C(1 + |u_0|_{H^2}^2), \quad \forall \: t \ge 0,
\end{equation}
\begin{equation} \label{H^3 energy bound}
\begin{aligned}
\int_t^{t+\epsilon} |u(\tau)|_{H^3}^2 \,\mathrm{d} \tau \le &
C|u_0|_{H^2}^2(1+ |u_0|_{L^2}^2) \\
 \vspace{2em}&+ \epsilon C(1+ |u_0|_{H^2}^2)^{10}, \quad \forall \: t \ge 0 \text{ and } 0< \epsilon <1.
\end{aligned}
\end{equation}

\el

\bp
Taking the time derivative of the free energy functional given in \eqref{new G} and using assumption ($\mathcal{H}$), we have
\begin{equation*}  
\begin{split}
\frac{\mathrm{d} \mathcal{G}}{\mathrm{d} t} =& \langle \frac{\delta \mathcal{G}}{\delta u},\, u_t \rangle = \langle \mu,\, \nabla \cdot (H(u)\nabla \mu) \rangle = -\int_\Omega H(u) |\nabla \mu|^2 \, \mathrm{d} x \le  0,
\end{split}
\end{equation*}
where $\mu$ is as in \eqref{new mu}. Thus,
\begin{equation} \label{G est.}
\begin{split}
\mathcal{G}(u(t)) \le &  \mathcal{G}(u(0)) = \int_\Omega
\biggl ( \frac{\alpha}{2}|\nabla u_0|^2 + \frac{1}{2}b_1u_0^2 +
\frac{1}{3}b_2u_0^3+\frac{1}{4}b_3u_0^4 \biggr) \,\mathrm{d} x\\
\le & \frac{\alpha}{2}|u_0|_{H^1}^2 + |u_0|_{L^\infty}^2 \int_\Omega (b_3 u_0^2 + C) \, \mathrm{d} x\\
\le & C|u_0|_{H^2}^2(1+|u_0|_{L^2}^2), \quad \forall \: t\ge 0,
\end{split}
\end{equation}
which justifies \eqref{G bound}.

By \eqref{new G} and \eqref{G est.}, we have
\begin{equation*}
\begin{split}
\int_\Omega \biggl ( \frac{\alpha}{2}|\nabla u|^2 + \frac{1}{2}b_1u^2 +
\frac{1}{3}b_2u^3+\frac{1}{4}b_3u^4 \biggr) \,\mathrm{d} x =
\mathcal{G}(u(t)) \le  & C|u_0|_{H^2}^2(1+|u_0|_{L^2}^2),
\end{split}
\end{equation*}
which implies
\begin{equation*}
\begin{split}
\alpha|\nabla u|^2 + \frac{1}{2} & b_3|u|_{L^4}^4 \le \int_\Omega \bigl(|b_1|u^2
+ \frac{2}{3}|b_2||u|^3 \bigr)\,\mathrm{d} x + C|u_0|_{H^2}^2(1+|u_0|_{L^2}^2) \\
\le & \int_\Omega \biggl ( \Bigl ( \frac{|b_1|^2}{b_3}+ \frac{1}{4}b_3u^4 \Bigr ) +
\Bigl (C\frac{b_2^4}{b_3^3}+ \frac{1}{4}b_3 |u|^4 \Bigr) \biggr )\,\mathrm{d} x +
C|u_0|_{H^2}^2(1+|u_0|_{L^2}^2) \\
=& \frac{1}{2}b_3|u|_{L^4}^4 + C|u_0|_{H^2}^2(1+|u_0|_{L^2}^2)+C.
\end{split}
\end{equation*}
We thus obtain
\begin{equation*}
\begin{split}
|\nabla u|^2 \le C|u_0|_{H^2}^2(1+|u_0|_{L^2}^2)+C< C(1+|u_0|_{H^2}^2)^2,
\end{split}
\end{equation*}
and \eqref{H^1 uniform bound} follows by the Poincar\'{e}'s inequality.

Recall that $\mu = -\alpha \Delta u + b_1u + b_2 u^2 + b_3u^3$. We have
by triangle inequality
\begin{equation*}
\begin{split}
\alpha |\nabla & \Delta u|^2 \le  2 |\nabla \mu|^2 + 2 \int_\Omega
|\nabla (b_1u +
b_2 u^2 + b_3u^3)|^2 \, \mathrm{d} x \\
\le & 2 |\nabla \mu|^2  + C (|\nabla u|^2 + |u|_{L^4}^2
\,|\nabla u|_{L^4}^2 + |u|_{L^8}^4\,|\nabla u|_{L^4}^2) \\
\le & 2 |\nabla \mu|^2  + C |u|_{H^1}^2 + C |u|_{H^1}^2\,
|u|_{H^1}^\frac{5}{4}\, |\nabla \Delta u|^\frac{3}{4} + C
|u|_{H^\frac{9}{8}}^4\, |u|_{H^1}^\frac{5}{4}\, |\nabla \Delta u|^\frac{3}{4} \\
\le & 2 |\nabla \mu|^2  + C |u|_{H^1}^2 + C |u|_{H^1}^\frac{13}{4}\,
|\nabla \Delta u|^\frac{3}{4} + C
|u|_{H^1}^\frac{15}{4}\, |\nabla \Delta u|^\frac{1}{4}\, |u|_{H^1}^\frac{5}{4}\, |\nabla \Delta u|^\frac{3}{4} \\
\le & 2 |\nabla \mu|^2  + C |u|_{H^1}^2 + C
|u|_{H^1}^{\frac{26}{5}}+ \frac{\alpha}{4} |\nabla \Delta u|^2 + C
|u|_{H^1}^{10}+ \frac{\alpha}{4}|\nabla \Delta u|^2,
\end{split}
\end{equation*}
where in the second last inequality we used the interpolation inequality $|u|_{H^\frac{9}{8}}^4 \le C |u|_{H^1}^\frac{15}{4}|u|_{H^3}^\frac{1}{4}$ and the fact that $|u|_{H^3}$ is equivalent to $|\nabla \Delta u|$ as shown in Lemma \ref{equi norm lemma}.  Then
\begin{equation} \label{H^3 est.}
\begin{split}
|\nabla \Delta u|^2 \le & C |\nabla \mu|^2  + C ( |u|_{H^1}^{10} + 1
).
\end{split}
\end{equation}
Note also
\begin{equation} \label{chemical potential est.}
\begin{split}
B_1 \int_t^{t+\epsilon} \int_\Omega |\nabla \mu|^2 \, \mathrm{d} x
\, \mathrm{d} \tau  \le & \int_t^{t+1} \int_\Omega H(u) |\nabla
\mu|^2
\, \mathrm{d} x \, \mathrm{d} \tau \\
= & \mathcal{G}(u(t))-\mathcal{G}(u(t+1)) \\
\le & C|u_0|_{H^2}^2(1+|u_0|_{L^2}^2).
\end{split}
\end{equation}
By \eqref{H^1 uniform bound}, \eqref{H^3
est.} and \eqref{chemical potential est.}, we have
\begin{equation} \label{H^3 est. 2}
\begin{split}
\int_t^{t+\epsilon} |\nabla \Delta u|^2 \, \mathrm{d} \tau
\le & C|u_0|_{H^2}^2(1+ |u_0|_{L^2}^2) +  \epsilon C
(1+|u_0|_{H^2}^2)^{10}.
\end{split}
\end{equation}

Now \eqref{H^3 energy bound} follows from \eqref{H^3 est. 2} and the fact that $|\nabla \Delta u|$ is an equivalent norm to $|u|_{H^3}$.

\ep

With the above two lemmas at our disposal,  we are ready to prove the following  local well-posedness result.
\begin{prop}  \label{local existence}

For any initial datum $u_0 \in W$,
there exist $T_0>0$ and a unique local solution $u(t)$ to the problem \eqref{CHE recall} such that:
\begin{equation} \label{weak formulation}
\begin{split}
& u\in L^2(0,\,T_0; \,W_1) \cap
C([0,\, T_0];\, W) \text{ with } \frac{\mathrm{d} u}{\mathrm{d} t} \in L^2(0,T_0; L^2(\Omega)), \\
& \fraction{\mathrm{d} }{\mathrm{d} t}\langle u,\, v \rangle =
\int_\Omega \bigl ( H(u)\mu \Delta v + H'(u)\mu \nabla u \cdot \nabla v \bigr) \,
\mathrm{d} x, \quad \forall \: v \in W \text{ and a.e. } 0\le t < T_0,  \\
& u(0)= u_0.
\end{split}
\end{equation}
\end{prop}

\bp
The proof consists of several steps.

{\sc Step 1.}
Given any $m \in \mathbb{N}$, let
\begin{equation} \label{finite app of W}
W_m = \text{span}\{e_k\:|\: 1\le k\le m\} \subset H^2, \quad
\widetilde{W}_m = C^1([0,\, T_m],\, W_m),
\end{equation}
where $e_k$'s are eigenvectors of $-\Delta$ with Neumann boundary
condition on $\partial \Omega$ and $\int_\Omega e_k \,\mathrm{d}x=0$, and $T_m>0$ is a constant to be chosen as follows.

According to standard existence theory for ordinary differential
equations, for each $m$, there exist $T_m>0$ and an approximate
solution $u_m$  to \eqref{weak formulation} in the
following sense:
\begin{equation} \label{Galerkin}
\begin{aligned}
&u_m=\sum^m_{j=0}x_j(t)e_j \in \widetilde{W}_m,\, x_j(t) \in \mathbb{R}, \\
& \fraction{\mathrm{d} }{\mathrm{d} t}\langle u_m,\, w \rangle =
\int_\Omega \bigl( H(u_m)\mu_m \Delta w + H'(u_m)\mu_m \nabla u_m \cdot
\nabla w \bigr) \,
\mathrm{d} x, \, \forall w \in W_m, \\
&u_m(0)=\sum_{j=1}^m \langle u_0, \, e_j \rangle e_j,
\end{aligned}
\end{equation}
where $\mu_m = -\alpha \Delta u_m + b_1u_m + b_2u_m^2 + b_3u_m^3$.

In order to show that there exists a solution to the original system, we need to establish some uniform estimates on the approximate solutions, which is the direction that we turn now.

In \eqref{Galerkin}, using $\Delta^2 u_m$ as the test
function, integration by parts twice and applying ($\mathcal{H}$), we obtain
\begin{equation} \label{App H4}
\begin{aligned}
\fraction{1}{2}\fraction{\mathrm{d}}{\mathrm{d} t} |\Delta
u_m|^2+\alpha B_1 |\Delta^2u_m|^2 & \le   \langle H(u_m) \Delta (b_1 u_m + b_2 u_m^2 + b_3 u_m^3),\,
\Delta^2 u_m\rangle \\
& \quad + \langle H'(u_m) \nabla u_m \cdot \nabla \mu_m,\, \Delta^2
u_m\rangle, \\
&:=I_1 + I_2.
\end{aligned}
\end{equation}
We have the following estimates for $I_1$ and $I_2$.
\begin{equation} \label{I_1}
\begin{aligned}
I_1=& \langle H(u_m) \Delta (b_1 u_m + b_2 u_m^2 + b_3 u_m^3),\,
\Delta^2 u_m\rangle \\
\le & C\int_\Omega (|u_m|^{p+1}+1)\bigl | \Delta (b_1 u_m + b_2 u_m^2 + b_3
u_m^3)\Delta^2 u_m \bigr |\, \mathrm{d} x.
\end{aligned}
\end{equation}
Note that
\begin{equation*}
\begin{aligned}
\int_\Omega\bigl | b_3 (|u_m|^{p+1}+1) & \Delta u_m^3 \Delta^2 u_m \bigr | \,
\mathrm{d} x \\
=& \int_\Omega \bigl |b_3 (|u_m|^{p+1}+1)( 3u_m^2 \Delta u_m + 6u_m |\nabla
u_m|^2) \Delta^2 u_m\bigr |\, \mathrm{d} x \\
\le & C(|u_m|_{L^\infty}^{p+1}+1)|u_m|_{L^\infty}^2|\Delta
u_m||\Delta^2 u_m| \\
& + C(|u_m|_{L^\infty}^{p+1}+1)|u_m|_{L^\infty}|\nabla
u_m|_{L^4}^2|\Delta^2 u_m| \\
\le & C(|u_m|_{H^2}^{p+1}+1)|u_m|_{H^2}^3|\Delta^2 u_m| \\
\le & \frac{\alpha B_1}{12}|\Delta^2 u_m|^2 +
C(|u_m|_{H^2}^{p+1}+1)^2 |u_m|_{H^2}^6.
\end{aligned}
\end{equation*}

Similarly, we have
\begin{equation*}
\begin{aligned}
\int_\Omega \bigl |b_1 (|u_m|^{p+1}+1) & \Delta u_m \Delta^2 u_m \bigr |\, \mathrm{d} x
\le  \frac{\alpha B_1}{12}|\Delta^2 u_m|^2 +
C(|u_m|_{H^2}^{p+1}+1)^2|u_m|_{H^2}^2,
\end{aligned}
\end{equation*}
and
\begin{equation*}
\begin{aligned}
\int_\Omega \bigl |b_2 (|u_m|^{p+1}+1) & \Delta u_m^2 \Delta^2 u_m \bigr |\,
\mathrm{d} x
\le  \frac{\alpha B_1}{12}|\Delta^2 u_m|^2 +
C(|u_m|_{H^2}^{p+1}+1)^2|u_m|_{H^2}^4.
\end{aligned}
\end{equation*}
Plugging the above three inequalities into \eqref{I_1}, we have
\begin{equation} \label{I_1 2}
\begin{aligned}
I_1 \le & \frac{\alpha B_1}{4}|\Delta^2 u_m|^2 +
C(|u_m|_{H^2}^{p+1}+1)^2(|u_m|_{H^2}^2 + |u_m|_{H^2}^4 + |u_m|_{H^2}^6) \\
\le & \frac{\alpha B_1}{4}|\Delta^2 u_m|^2 +
C(|u_m|_{H^2}^2+1)^{p+4}.
\end{aligned}
\end{equation}
For $I_2$ , we have
\begin{equation*}
\begin{aligned}
I_2 =& \langle H'(u_m) \nabla u_m \cdot \nabla \mu_m,\, \Delta^2 u_m\rangle \\
=& \langle -\alpha H'(u_m) \nabla u_m \cdot \nabla \Delta u_m,\, \Delta^2 u_m\rangle\\
& + \langle H'(u_m) \nabla u_m \cdot \nabla (b_1 u_m + b_2 u_m^2 + b_3 u_m^3),\, \Delta^2 u_m\rangle.
\end{aligned}
\end{equation*}
By our assumption on $\mathcal{H}$, the first part of $I_2$ can be estimated as
\begin{equation*}
\begin{aligned}
\langle -\alpha H'(u_m) \nabla u_m \cdot & \nabla \Delta u_m,\, \Delta^2 u_m\rangle \\
\le & C (|u_m|^p_{L^\infty}+1) |\nabla u_m|_{L^\infty} |\nabla \Delta u_m| |\Delta^2 u_m| \\
\le & \mbox{(by Agmon's inequality, see e.g. \cite{Tem97} page 52)} \\
\le & C (|u_m|^p_{H^2}+1) |\nabla u_m|_{H^1}^{\frac{1}{2}}  |\nabla u_m|_{H^2}^{\frac{1}{2}} |u_m|_{H^3} |\Delta^2 u_m| \\
\le & C (|u_m|^p_{H^2}+1) |u_m|_{H^2}^{\frac{1}{2}} |u_m|_{H^3}^{\frac{3}{2}} |\Delta^2 u_m| \\
\le & C (|u_m|^p_{H^2}+1) |u_m|_{H^2}^{\frac{5}{4}} |\Delta^2 u_m|^{\frac{7}{4}} \\
\le &  \frac{\alpha B_1}{8}|\Delta^2 u_m|^2 +
C(|u_m|_{H^2}^p+1)^8|u_m|_{H^2}^{10}.
\end{aligned}
\end{equation*}
The second part of $I_2$ can be estimated in the same fashion, and we have
\begin{equation*}
\begin{aligned}
 \langle H'(u_m) \nabla u_m \cdot & \nabla (b_1 u_m + b_2 u_m^2 + b_3 u_m^3),\, \Delta^2 u_m\rangle \\
 \le &  \frac{\alpha B_1}{8}|\Delta^2 u_m|^2 +
C(|u_m|_{H^2}^p+1)^2( |u_m|_{H^2}^4 + |u_m|_{H^2}^6+ |u_m|_{H^2}^8).
\end{aligned}
\end{equation*}
By combining the estimates for the two parts of $I_2$, we obtain
\begin{equation} \label{I_2}
\begin{aligned}
I_2 \le & \frac{\alpha B_1}{4}|\Delta^2 u_m|^2 + C(|u_m|_{H^2}^p+1)^2( |u_m|_{H^2}^4+  |u_m|_{H^2}^6+ |u_m|_{H^2}^8)  \\
& \hspace{8em}+   C(|u_m|_{H^2}^p+1)^8 |u_m|_{H^2}^{10} \\
\le & \frac{\alpha B_1}{4}|\Delta^2 u_m|^2 + C(|u_m|_{H^2}^2+1)^{4p+5}.
\end{aligned}
\end{equation}

By \eqref{App H4}, \eqref{I_1 2}, \eqref{I_2} and Lemma \ref{equi norm lemma}, we have
\begin{equation} \label{energy bound local existence}
\begin{aligned}
\fraction{\mathrm{d}}{\mathrm{d} t} |\Delta u_m|^2+ \alpha B_1
|\Delta^2u_m|^2 \le & C(u_0)(|u_m|_{H^2}^2+1)^{4p+5} \\
 \le & C(u_0)(|\Delta u_m|^2+1)^{4p+5}.
\end{aligned}
\end{equation}

Set $$y=1+|\Delta u_m|^2,$$
then by \eqref{energy bound local existence}
\begin{equation}
\begin{aligned}
\fraction{\mathrm{d}y}{\mathrm{d} t}  \le & C(u_0)y^{4p+5}.
\end{aligned}
\end{equation}
Integrating this differential inequality, we find
$$0< y(t) \le \bigl (y(0)^{-4(p+1)} - C(u_0) t  \bigr) ^{-1/(4p+4 )}$$
for $0\le t \leq T_0$ where
\[ 0<T_0< \frac{1}{C(u_0)(1+ |\Delta u_0|^2)^{4(p+1)}}.\]

This together with \eqref{energy bound local existence} implies that $T_m$ as in \eqref{finite app of W} satisfies $T_m \ge T_0$ for each $m$ and
\begin{equation}  \label{spatial regularity}
\begin{aligned}
u_m \in \mbox{ a bounded set of } L^2(0,\,T_0; W_1) \cap
L^\infty(0,\,T_0;W) ,
\end{aligned}
\end{equation}
independent of $m$.

Now, by \eqref{CHE recall} and \eqref{spatial regularity} we have the following estimate for $|\frac{\mathrm{d}
u_m}{\mathrm{d} t}|_{L ^2(0,\,T_0; \,L^2)}$:
\begin{equation} \label{time regularity est.}
\begin{split}
|\frac{\mathrm{d} u_m}{\mathrm{d} t}|_{L^2(0,\, T_0; \,L^2)}^2 = &
\int_0^T \int_\Omega \left|\nabla \cdot \left[ H(u_m) \nabla \mu_m
\right]\right|^2 \, \mathrm{d}
x \, \mathrm{d} t \\
\le & C(u_0) \bigl ( |\Delta^2 u_m|_{L^2(0,\, T_0; \,L^2)}^2 + |\nabla u_m \cdot
\nabla
\Delta u_m|_{L^2(0,\, T_0; \,L^2)}^2 \\
&+ |\Delta (b_1 u_m + b_2 u_m^2 + b_3 u_m^3)|_{L^2(0,\, T_0; \,L^2)}^2 \\
&+ |\nabla u_m \cdot \nabla (b_1 u_m + b_2 u_m^2 + b_3
u_m^3)|_{L^2(0,\, T_0; \,L^2)}^2 \bigr).
\end{split}
\end{equation}

Note that
\begin{equation*}
\begin{split}
|\nabla u_m \cdot \nabla \Delta u_m|_{L^2} \le
|u_m|_{H^1}|u_m|_{H^3} \le &
C |u_m|_{H^1}|u_m|_{H^4} \\
\le & C(|u_m|_{H^1}^2+|u_m|_{H^4}^2),
\end{split}
\end{equation*}
which together with \eqref{H^1 uniform bound} implies
\begin{equation*}
\begin{split}
|\nabla u_m \cdot \nabla \Delta u_m|_{L^2(0,\, T_0; \,L^2)}^2 \le
C(C(u_0)T_0 +|\Delta ^2u_m|_{L^2(0,\, T_0; \,L^2)}^2).
\end{split}
\end{equation*}
Similarly, we have
\begin{equation*}
\begin{split}
|\Delta (b_1 u_m + b_2 u_m^2 + b_3 u_m^3)|_{L^2(0,\, T_0; \,L^2)}^2 \le
C(C(u_0)T_0 +|\Delta ^2u_m|_{L^2(0,\, T_0; \,L^2)}^2),
\end{split}
\end{equation*}
and
\begin{equation*}
\begin{split}
|\nabla u_m \cdot \nabla (b_1 u_m + b_2 u_m^2 + b_3
u_m^3)|_{L^2(0,\, T_0; \,L^2)}^2 \le C(C(u_0)T_0 +|\Delta
^2u_m|_{L^2(0,\, T_0; \,L^2)}^2).
\end{split}
\end{equation*}
Plugging the above four inequalities in \eqref{time regularity est.}, we obtain
\begin{equation*}
\begin{split}
|\frac{\mathrm{d} u_m}{\mathrm{d} t}|_{L^2(0,\,T_0; \,L^2)}^2 \le &
C(u_0)(T_0 +|\Delta ^2u_m|_{L^2(0,\,T_0; \,L^2)}^2) ,
\end{split}
\end{equation*}
which together with \eqref{spatial regularity} shows that
\begin{equation} \label{time regularity}
\begin{split}
\frac{\mathrm{d} u_m}{\mathrm{d} t} \in \mbox{ a bounded set of } L^2 (0,\,T_0; \,L^2).
\end{split}
\end{equation}

{\sc Step 2.}
By \eqref{spatial regularity} and \eqref{time regularity} we can extract a subsequence $u_{m'}$ of $u_m$ which satisfies
\begin{equation} \label{weak conv. 1}
\begin{split}
& u_{m'} \rightharpoonup u \mbox{ weakly in } L^2 (0, T_0; W_1), \\
& u_{m'} \rightharpoonup u \mbox{ weak-star in } L^\infty (0, T_0; W), \\
& \frac{\mathrm{d} u_{m'}}{\mathrm{d} t} \rightharpoonup \frac{\mathrm{d} u}{\mathrm{d} t} \mbox{ weakly in } L^2 (0, T_0; L^2).
\end{split}
\end{equation}
Thanks to the compactness of the embedding of $W_1$ in $H^3\cap W$, the inclusion
$$\{ f \in L^2 (0,T_0; W_1) \mid  \frac{\mathrm{d} f}{\mathrm{d} t} \in L^2 (0,T_0; L^2)\} \subset  L^2 (0,T_0; W\cap H^3)$$
is compact; see e.g. \cite{Boy99} Lemma 1.6. Therefore without loss of generality, we may assume
\begin{equation} \label{weak conv. 2}
\begin{split}
u_{m'} \rightarrow u \mbox{ strongly in } L^2 (0, T_0; W\cap H^3).
\end{split}
\end{equation}
Note also the following embedding is continuous
\begin{equation*}
\{\, f\in L^2  (0,\,  T_0; \, W_1),\, \frac{\mathrm{d} f}{\mathrm{d} t}
\in L^2  (0,\,  T_0; \, L^2)\,\} \hookrightarrow C([0,\,T_0]; W),
\end{equation*}
see e.g. \cite{Boy99}.
Thus, upon passing to a further subsequence, we have by \eqref{weak conv. 1}$_1$ and \eqref{weak conv. 1}$_3$
\begin{equation}  \label{weak conv. 3}
\begin{split}
 u_{m'} \rightharpoonup u \mbox{ weakly in } C([0,T_0]; W).
\end{split}
\end{equation}
In particular, $u_{m'}(0)$ converges  weakly to $u(0)$ in $W$, and so $u(0)=u_0$ because $u_{m'}(0)$ converges to $u_0$ strongly in $W$. We still need to show that the function $u$ satisfies \eqref{weak formulation}$_2$.

We consider $\phi \in C^\infty_c(0,T_0)$ and $N\ge 1$. For any $m' \ge N$, $u_{m'}$ satsfies \eqref{Galerkin}$_2$ with $w = e_N$ where $e_N$  is as in \eqref{finite app of W}. We multiply this equation by $\phi(t)$ and integrate by parts to obtain
\begin{equation}  \label{pass to the limit}
\begin{split}
- \int_0^{T_0} \int_\Omega u_{m'} & e_N \phi' \, \mathrm{d}x \, \mathrm{d}t  \\
&= \int_0^{T_0} \int_\Omega \bigl( H(u_{m'})\mu_{m'} \Delta e_N + H'(u_{m'})\mu_{m'} \nabla u_{m'} \cdot \nabla e_N \bigr)  \phi  \, \mathrm{d} x \, \mathrm{d} t.
\end{split}
\end{equation}
The convergence properties of the sequence $u_{m'}$ allow us to pass to the limit in this equation. The passage to the limit on the LHS is easy to see by using \eqref{weak conv. 1}$_1$, and we have
\begin{equation}
\begin{split}
\int_0^{T_0} \int_\Omega u_{m'} e_N \phi' \, \mathrm{d}x \, \mathrm{d}t  \stackrel{m' \to \infty}{\xrightarrow{\hspace*{2em}}} \int_0^{T_0} \int_\Omega u e_N \phi' \, \mathrm{d}x \, \mathrm{d}t.
\end{split}
\end{equation}
For the RHS, we have
\begin{equation}
\begin{split}
&\int_0^{T_0} \int_\Omega \bigl( H(u_{m'})\mu_{m'} \Delta e_N + H'(u_{m'})\mu_{m'} \nabla u_{m'} \cdot \nabla e_N \bigr)  \phi  \, \mathrm{d} x \, \mathrm{d} t \\
&\hspace{1em}=  -\alpha \int_0^{T_0} \int_\Omega  H(u_{m'}) \Delta u_{m'} \Delta e_N \phi \, \mathrm{d} x \, \mathrm{d} t \\
&\hspace{1em} +  \int_0^{T_0} \int_\Omega  H(u_{m'}) (b_1u_{m'} + b_2u_{m'}^2 + b_3u_{m'}^3 ) \Delta e_N \phi \, \mathrm{d} x \, \mathrm{d} t \\
&\hspace{1em} +  \int_0^{T_0} \int_\Omega  H'(u_{m'})(-\alpha \Delta u_{m'} +b_1u_{m'} + b_2u_{m'}^2 + b_3u_{m'}^3) \nabla u_{m'} \cdot \nabla e_N  \phi  \, \mathrm{d} x \, \mathrm{d} t.
\end{split}
\end{equation}
For brevity, we will only show the convergence of the first term and the convergence of the rest terms follows in the same fashion.
\begin{equation}  \label{1st part}
\begin{split}
& \bigl |  \int_0^{T_0} \int_\Omega  H(u_{m'}) \Delta u_{m'} \Delta e_N \phi \, \mathrm{d} x \, \mathrm{d} t - \int_0^{T_0} \int_\Omega  H(u) \Delta u \Delta e_N \phi \, \mathrm{d} x \, \mathrm{d} t  \bigr |  \\
& \hspace{3em} \le  \bigl |  \int_0^{T_0} \int_\Omega  \bigl ( H(u_{m'}) - H(u) \bigr ) \Delta u \Delta e_N \phi \, \mathrm{d} x \, \mathrm{d} t \bigr |  \\
& \hspace{5em}+ \bigl |  \int_0^{T_0} \int_\Omega  H(u) (\Delta u_{m'} - \Delta u) \Delta e_N \phi \, \mathrm{d} x \, \mathrm{d} t \bigr | .
\end{split}
\end{equation}
Using \eqref{weak conv. 1} -- \eqref{weak conv. 3} and mean value theorem, the first quantity on the RHS of \eqref{1st part} can be estimated as
\begin{equation}
\begin{split}
\bigl |  \int_0^{T_0} & \int_\Omega  \bigl ( H(u_{m'}) - H(u) \bigr ) \Delta u \Delta e_N \phi \, \mathrm{d} x \, \mathrm{d} t \bigr | \\
 & \le \bigl |  \int_0^{T_0} \int_\Omega  H'(w_{m'})(u_{m'} - u)  \Delta u \Delta e_N \phi \, \mathrm{d} x \, \mathrm{d} t \bigr | \\
& \le C \int_0^{T_0} \int_\Omega  (|w_{m'}|^p + 1) |u_{m'} - u|  | \Delta u| |\Delta e_N|| \phi| \, \mathrm{d} x \, \mathrm{d} t  \\
& \le C (|w_{m'}|^p_{L^{\infty}(0,  T_0; W)} + 1) |u|_{L^{\infty}(0,  T_0; W)} |\phi|_{L^\infty} \int_0^{T_0} \int_\Omega   |u_{m'} - u|  |\Delta e_N| \, \mathrm{d} x \, \mathrm{d} t  \\
& \le C (|w_{m'}|^p_{L^{\infty}(0,  T_0; W)} + 1) |u|_{L^{\infty}(0,  T_0; W)} |\phi|_{L^\infty} \int_0^{T_0} |u_{m'} - u|_{L^2}|e_N|_{H^2}\, \mathrm{d} t \\
& \le C (|w_{m'}|^p_{L^{\infty}(0,  T_0; W)} + 1) |u|_{L^{\infty}(0,  T_0; W)} |\phi|_{L^\infty} \int_0^{T_0} (\frac{1}{\delta} |u_{m'} - u|^2_{L^2} + \delta |e_N|^2_{H^2})\, \mathrm{d} t \\
& \le C \delta (|w_{m'}|^p_{L^{\infty}(0,  T_0; W)} + 1) |u|_{L^{\infty}(0,  T_0; W)} |\phi|_{L^\infty} |e_N|^2_{L^{\infty}(0,  T_0; W)}  \\
&\hspace{2em} + \frac{C}{\delta}(|w_{m'}|^p_{L^{\infty}(0,  T_0; W)} + 1) |u|_{L^{\infty}(0,  T_0; W)} |\phi|_{L^\infty} |u_{m'} - u|^2_{L^2(0,  T_0; W)}.
\end{split}
\end{equation}
In light of \eqref{weak conv. 2}, the above quantity can be made as small as possible by choosing $\delta>0$ sufficiently small and $m'$ sufficiently large. The second quantity on the RHS of \eqref{1st part} can be estimated in the same way.

Now, we obtain after passing to the limit the following equation for $u$:
\begin{equation*}
\begin{split}
- \int_0^{T_0} \int_\Omega u & e_N \phi' \, \mathrm{d}x \, \mathrm{d}t  = \int_0^{T_0} \int_\Omega \bigl( H(u)\mu \Delta e_N + H'(u)\mu \nabla u \cdot \nabla e_N \bigr)  \phi  \, \mathrm{d} x \, \mathrm{d} t.
\end{split}
\end{equation*}
The limit equation obtained above is fulfilled for any $N$ and any $\phi \in C^\infty_c(0,T_0)$, so that the density of $span\{e_N \mid N\in \mathbb{N}\}$ in $W$ allows us to conclude that $u$ satisfies \eqref{weak formulation}$_2$.

{\sc Step 3.} In the following, we will sketch the proof for the uniqueness. Let $u_1$ and $u_2$ be any two strong solutions of \eqref{CHE
recall} defined on the interval $[0,\, T_0]$. There exists $C(T_0)>0$ such that for $i=1,2$
\begin{equation} \label{uniform bounds}
|u_i(t)|_{H^2} \le  C(T_0), \quad \forall t \in [0,\, T_0].
\end{equation}

Let $\tilde{u}=u_1 - u_2$. Multiplying \eqref{CHE recall} by $v\in H^1$, integrating over $\Omega$, we get:
\begin{equation} \label{weak formulation 2}
\begin{aligned}
\langle \fraction{\mathrm{d} u}{\mathrm{d} t},\, v \rangle =&
\langle H(u)\nabla (\alpha \Delta u - (b_1u+b_2u^2 +
b_3u^3)),\, \nabla v \rangle,
\end{aligned}
\end{equation}
from which we see that $\tilde{u}$ satisfies
\begin{equation}
\begin{aligned}
\langle \fraction{\mathrm{d} \tilde{u}}{\mathrm{d} t},\, v \rangle
=& \langle \alpha H(u_1)\nabla \Delta \tilde{u},\, \nabla v \rangle
+
\langle \alpha ( H(u_1)- H(u_2)) \nabla \Delta u_2,\, \nabla v \rangle \\
&- \langle H(u_1) (b_1u_1+b_2u_1^2 + b_3u_1^3 - b_1u_2-b_2u_2^2-
b_3u_2^3), \, \nabla v \rangle \\
&-\langle ( H(u_1)- H(u_2)) (b_1u_2+b_2u_2^2 + b_3u_2^3),\, \nabla v
\rangle.
\end{aligned}
\end{equation}
From the regularity we obtained for solutions of \eqref{CHE recall},
we can take $v$ in the above equation to be $-\Delta \tilde{u}$ and
use ($\mathcal{H}$) to get:
\begin{equation} \label{difference est.}
\begin{aligned}
\frac{1}{2}\frac{\mathrm{d} |\nabla \tilde{u}|^2}{\mathrm{d} t} + &
\alpha B_1 |\nabla \Delta \tilde{u}|^2 \le
 - \langle \alpha ( H(u_1)- H(u_2)) \nabla \Delta u_2,\, \nabla \Delta \tilde{u} \rangle \\
&+ \langle H(u_1) (b_1u_1+b_2u_1^2 + b_3u_1^3 - b_1u_2-b_2u_2^2-
b_3u_2^3), \, \nabla \Delta \tilde{u} \rangle \\
&+\langle ( H(u_1)- H(u_2)) (b_1u_2+b_2u_2^2 + b_3u_2^3),\, \nabla
\Delta \tilde{u} \rangle.
\end{aligned}
\end{equation}
Here, the term $\frac{\mathrm{d} |\nabla \tilde{u}|^2}{\mathrm{d}t}$ is understood in the distribution sense. More specifically, since $\frac{\mathrm{d} \tilde{u}}{\mathrm{d}t} \in L^2(0,\, T_0; L^2)$ and $\tilde{u} \in L^2(0,\,T_0;W_1)$, then by Theorem 2.3 in \cite{LM72}, we know that $\frac{\mathrm{d} u}{\mathrm{d}t} \in L^2(0,T_0; W)$.

Denote the terms on the RHS of \eqref{difference est.} by $I_3,\,
I_4,\, I_5$. We have the following estimates for them.

Applying mean value theorem to $H$ and using \eqref{uniform bounds}, we have
\begin{equation*}
\begin{aligned}
I_3 =& - \langle \alpha ( H(u_1)- H(u_2)) \nabla \Delta u_2,\, \nabla \Delta \tilde{u} \rangle \\
\le & C |H'(w)|_{L^\infty} |\tilde{u}|_{L^\infty} \, |\nabla \Delta u_2| \, |\nabla \Delta \tilde{u}| \\
\le & \mbox{(by Agmon's inequality)} \\
\le & C (1+|w|_{L^\infty}^p) |\tilde{u}|_{H^1}^\frac{1}{2}\,
|\tilde{u}|_{H^2}^\frac{1}{2} \, |\nabla \Delta u_2| \, |\nabla
\Delta \tilde{u}| \\
\le & C (1+|w|_{L^\infty}^p) |\tilde{u}|_{H^1}^\frac{3}{4}\,
|\tilde{u}|_{H^3}^\frac{1}{4} \, |\nabla \Delta u_2| \, |\nabla
\Delta \tilde{u}| \\
\le & C(1+|w|_{H^2}^p)^\frac{8}{3} |\nabla \Delta u_2|^\frac{8}{3}\,
|\tilde{u}|_{H^1}^2 +
\frac{\alpha B_1}{3} |\nabla \Delta \tilde{u}|^2 \\
\le & C(1+C(T_0)^p)^\frac{8}{3} (|u_2|_{H^4}^2+1) |\tilde{u}|_{H^1}^2 +
\frac{\alpha B_1}{3} |\nabla \Delta \tilde{u}|^2,
\end{aligned}
\end{equation*}
where $w=\theta(t)u_1 + (1-\theta(t))u_2$ for some $\theta(t)$.

By \eqref{uniform bounds}, we have
\begin{equation*}
\begin{aligned}
I_4= & \langle H(u_1) (b_1u_1+b_2u_1^2 + b_3u_1^3 - b_1u_2-b_2u_2^2-
b_3u_2^3), \, \nabla \Delta \tilde{u} \rangle \\
\le & C (1+|u_1|_{L^\infty}^{p+1}) (1+|u_1|_{L^\infty}^2+|u_2|_{L^\infty}^2)|\tilde{u}| |\nabla \Delta \tilde{u}| \\
\le & C(1+C(T_0)^{p+3})^2 |\tilde{u}|^2 + \frac{\alpha B_1}{3} |\nabla \Delta \tilde{u}|^2 \\
\le & C(1+C(T_0)^{p+3})^2 |\nabla \tilde{u}|^2 + \frac{\alpha B_1}{3}
|\nabla \Delta \tilde{u}|^2.
\end{aligned}
\end{equation*}
Similarly,
\begin{equation*}
\begin{aligned}
I_5= &\langle ( H(u_1)- H(u_2)) (b_1u_2+b_2u_2^2 + b_3u_2^3),\,
\nabla \Delta \tilde{u} \rangle \\
\le & C (1+C(T_0)^{p+3})|\tilde{u}| |\nabla \Delta \tilde{u}| \\
\le & C (1+C(T_0)^{p+3})^2 |\nabla \tilde{u}|^2 + \frac{\alpha B_1}{3}
|\nabla \Delta \tilde{u}|^2.
\end{aligned}
\end{equation*}
Plugging the above estimates in \eqref{difference est.}, we have
\begin{equation}
\begin{aligned}
\frac{\mathrm{d} |\nabla \tilde{u}|^2}{\mathrm{d} t} \le  C
(|u_2|_{H^4}^2+1) |\nabla \tilde{u}|^2,
\end{aligned}
\end{equation}
which together with $u\in L^2(0,\, T_0;\,W_1)$ and $|\nabla
\tilde{u}(0)|^2 = 0$ implies $|\nabla \tilde{u}(t)|^2 = 0$ for all
$t\in [0,\,T_0]$, and the uniqueness is thus proven.

\ep

\bp[{\bf \emph{Completion of the proof of Theorem \ref{Global well-posedness}}}]
For $1<p<3$, one can find $2<q_1,q_2<3$ such that the following inequalities are satisfied:
\begin{equation}
\begin{aligned}
& 6p < \frac{3q_1}{3-q_1}, \qquad  p\left(\frac{3}{2}-\frac{3}{q_1}\right) < 1, \\
& 3(p+3) < \frac{3q_2}{3-q_2}, \qquad  (p+3)\left(\frac{3}{2}-\frac{3}{q_2}\right) < 2.
\label{growth.cond}
\end{aligned}
\end{equation}

Taking $L^2$ inner product on both sides of \eqref{CHE recall} with
$\Delta^2 u$, we get:
\begin{equation} \label{time derivative of H2 norm}
\begin{split}
\frac{1}{2}\frac{\mathrm{d} }{\mathrm{d} t} |\Delta u|^2 =& -\langle
\alpha H(u)\Delta^2 u,\, \Delta^2 u \rangle - \langle \alpha H'(u)
\nabla u \cdot \nabla
\Delta u,\, \Delta^2 u \rangle \\
& + \langle H(u)\Delta (b_1 u+ b_2 u^2 +
b_3 u^3),\, \Delta^2 u \rangle \\
&+ \langle H'(u)\nabla u \cdot \nabla (b_1 u + b_2 u^2 + b_3 u^3),
\, \Delta^2 u \rangle.
\end{split}
\end{equation}
Here $\frac{1}{2}\frac{\mathrm{d}}{\mathrm{d} t} |\Delta u|^2$ on the LHS is understood in the scalar distribution sense on $(0,T)$; again see Theorem 2.3 in \cite{LM72}.
Then by our assumption ($\mathcal{H}$), we have
\begin{equation}  \label{scaled H2 norm}
\begin{split}
\frac{1}{2}\frac{\mathrm{d} }{\mathrm{d} t} |\Delta u|^2 +\alpha
B_1 |\Delta^2 u|^2 & \le  - \langle \alpha H'(u) \nabla u \cdot \nabla
\Delta u,\, \Delta^2 u \rangle \\
& \quad+ \langle H(u)\Delta (b_1 u+ b_2 u^2 +
b_3u^3),\, \Delta^2 u \rangle \\
&\quad + \langle H'(u)\nabla u \cdot \nabla (b_1 u + b_2u^2 + b_3u^3), \,
\Delta^2 u \rangle, \\
&:= J_1 + J_2 + J_3.
\end{split}
\end{equation}

Let $p_1=6p \frac{(3+\beta)}{(3-\beta)}$. Then for $\beta>0$
sufficiently small we have $p_1<\frac{3q_1}{3-q_1}$ by
\eqref{growth.cond}. Hence $W^{1,q_1}\hookrightarrow L^{p_1}$ and we
have
\begin{equation} \label{est.for.J_1}
|u|_{L^{p_1}}\leq C |\nabla u |_{L^{q_1}}\leq C |\nabla
u|^{\frac{3}{q_1}-\frac{1}{2}}|\nabla u|^{\frac{3}{2}-\frac{3}{q_1}}_{L^6}\leq C(u_0)
|u|^{\frac{3}{2}-\frac{3}{q_1}}_{H^2}.
\end{equation}
Using \eqref{est.for.J_1} and \eqref{growth.cond} we can obtain:
\begin{equation}
\begin{split}
|H'(u)\nabla u|_{L^{3+\beta}} & \leq  C |\nabla u|_{L^6} (1+|u|^p_{ L^{p_1}}) \leq C(u_0) |u|_{H^2}(1+|u|^{p(\frac{3}{2}-\frac{3}{q_1})}_{H^2}) \\
& \leq C(u_0) (1+|u|^{1+p(\frac{3}{2}-\frac{3}{q_1})}_{H^2} )  \\
&\leq C(u_0)(1+|u|^2_{H^2} ).
\end{split}
\label{est2.for.J_1}
\end{equation}
To estimate $J_1$, let $\eta$ be defined as
\begin{equation} \label{eta1}
\frac{1}{3+\beta}+\frac{1}{6-\eta}=\frac{1}{2}.
\end{equation}
and note that
\begin{equation}\label{est3.for.J_1}
|\nabla \Delta u|_{L^{6-\eta}}^{6-\eta}\leq |\nabla \Delta u|^{\eta/2}|\nabla \Delta u|_{L^6}^{6-3\eta/2}
\leq |u|_{H^1}^{\eta/6}|u|_{H^4}^{6-7\eta/6}.
\end{equation}
We estimate $J_1$ using \eqref{est2.for.J_1}, \eqref{eta1} and \eqref{est3.for.J_1} as follows:
\begin{equation} \label{J_1}
\begin{split}
J_1 =& - \langle \alpha H'(u) \nabla u \cdot \nabla \Delta u,\, \Delta^2 u \rangle \\
 \le & C |H'(u)\nabla u|_{L^{3+\beta}} |\nabla \Delta u|_{L^{6-\eta}} |\Delta^2 u| \\
\le & C(u_0) \, (1+|u|^2_{H^2})  \,|u|_{H^4}^{2-\eta/(36-6\eta)} \\
\le & C(u_0) \, (1+|u|^2_{H^2}) \,
(\epsilon^{-(12(6-\eta)-\eta)/\eta)}+\epsilon |u|_{H^4}^2).
\end{split}
\end{equation}
By \eqref{growth.cond}, we know $W^{1,q_2}\hookrightarrow^{3(p+3)}$, then
\begin{equation} \label{est.for.J_2}
|u|_{L^{3(p+3)}}\leq  C |\nabla u|_{L^{q_2}} \leq C |\nabla u|^{\frac{3}{q_2}-\frac{1}{2}}|\nabla u|^{\frac{3}{2}-\frac{3}{q_2}}_{L^6}
\leq  C(u_0) (|u|^{2/(p+3)}_{H^2}+1).
\end{equation}
To estimate $J_2$ we first estimate the following two integrals
\begin{equation} \label{J_2part1}
\begin{split}
 \int_{\Omega} (1+|u|^{p+3})|\Delta u| \, |\Delta^2 u| \, \mathrm{d}x \le & C  (1+|u|_{L^{3(p+3)}}^{p+3})|\Delta u|_{L^6}\, |\Delta^2 u| \\
\le & \text{by \eqref{est.for.J_2}}\\
\le & C(u_0)  (1+|u|^2_{H^2})\,|u|_{H^3}\, |u|_{H^4}\\
\le & C(u_0)  (1+|u|^2_{H^2})\,|u|^{1/3}_{H^1}\, |u|^{5/3}_{H^4}\\
\le &C(u_0)  (1+|u|^2_{H^2})\, (\epsilon^{-5}+\epsilon|u|^2_{H^4}).
\end{split}
\end{equation}
\begin{equation} \label{J_2part2}
\begin{split}
\int_\Omega (1+ |u|^{p+2}) |\nabla u|^2 |\Delta^2 u|\, \mathrm{d}x \le &
C \int_\Omega (1+ |u|^{p+3}) |\nabla u|^2 |\Delta^2 u| \\
\le & C (1+ |u|_{L^{3(p+3)}}^{p+3}) |\nabla u|_{L^{12}}^2 |\Delta^2 u| \\
\le & C(u_0) (1+ |u|^2_{H^2})  |u|_{H^{9/4}}^2 |u|_{H^4} \\
\le & C(u_0) (1+ |u|^2_{H^2})  |u|_{H^1}^{7/6}  |u|^{11/6}_{H^4} \\
\le & C(u_0)  (1+|u|^2_{H^2}) (\epsilon^{-11}+\epsilon |u|^2_{H^4}).
\end{split}
\end{equation}
Using \eqref{J_2part1} and \eqref{J_2part2} we have
\begin{equation} \label{J_2}
\begin{split}
J_2 &=\langle H(u) \Delta (b_1 u+b_2 u^2+ b_3 u^3),\, \Delta^2 u \rangle \\
\le  & C \int_{\Omega}(1+|u|^{p+1}) \left[(1+|u|^2)|\Delta
u|+(1+|u|)|\nabla u|^2 \right]\, |\Delta^2
u| \, \mathrm{d}x\\
\le & C(u_0)  (1+|u|^2_{H^2}) (\epsilon^{-11}+\epsilon |u|^2_{H^4}).
\end{split}
\end{equation}
For $J_3$, we have
\begin{equation} \label{J_3}
\begin{split}
J_3 =& \langle H'(u)\nabla u \cdot \nabla(b_1 u+b_2 u^2+ b_3 u^3), \, \Delta^2 u \rangle \\
\le & \int_\Omega (1+|u|^p)(1+|u|^2)|\nabla u|^2|\Delta^2 u| \, \mathrm{d}x\\
\le & \text{(by \eqref{J_2part2} )} \\
\le & C(u_0)  (1+|u|^2_{H^2}) (\epsilon^{-11}+\epsilon |u|^2_{H^4}).
\end{split}
\end{equation}
From the estimates for $J_1,\, J_2$ and $J_3$ given in \eqref{J_1},
\eqref{J_2} and \eqref{J_3} respectively, we have by \eqref{A(t)
notation} and \eqref{scaled H2 norm}:
\begin{equation}  \label{H^2 energy ineq}
\begin{split}
\frac{\mathrm{d} }{\mathrm{d} t} \mathcal{A}(t) +&\left(2 \alpha
B_1- C(u_0)\epsilon \mathcal{A}(t) \right) |\Delta^2 u|^2 \le
C(u_0)\epsilon^{-N} \mathcal{A}(t).
\end{split}
\end{equation}
Here $N=max\{11, (12(6-\eta)-\eta)/\eta\}$ with $\eta$ determined by
\eqref{eta1}. Also note that $N \rightarrow \infty$
as $p$ approaches the critical exponent $3$.

The crucial step towards the global existence and uniqueness is a
uniform $H^2$ bound for the solution. This can be achieved by
manipulating \eqref{H^2 energy ineq} when the initial data is small
as we now show.  To our knowledge, a similar method first
appeared in \cite{LinLiu95}.

First, for any $t\ge 0$ and $1>\tilde{\epsilon}>0$ to be specified
later, we have by \eqref{H^1 uniform bound} and \eqref{H^3 energy bound}:
\begin{equation} \label{A(t) est.}
\begin{split}
\int_t^{t+\tilde{\epsilon}} \mathcal{A}(\tau) \, \mathrm{d} \tau =&
\int_t^{t+\tilde{\epsilon}} (|u|_{H^2}^2 + 1)\, \mathrm{d} \tau \le
\int_t^{t+\tilde{\epsilon}} ( C|u|_{H^1}\,|u|_{H^3}+1 )\, \mathrm{d} \tau \\
\le & \int_t^{t+\tilde{\epsilon}} ( C|u|_{H^1}^2 + C|u|_{H^3}^2 + 1 )\, \mathrm{d} \tau  \\
\le & C \tilde{\epsilon} (1+ |u_0|_{H^2}^2)^2 + C|u_0|_{H^2}^2(1+ |u_0|_{L^2}^2) \\
& + C \tilde{\epsilon} (1 + |u_0|_{H^2}^2)^{10} + \tilde{\epsilon}.
\end{split}
\end{equation}

From now on we will assume that $|u_0|_{H^2} \le 1$. Then we have by \eqref{A(t) est.}
\begin{equation} \label{A(t) est. 2}
\begin{split}
\int_t^{t+\tilde{\epsilon}} \mathcal{A}(\tau) \, \mathrm{d} \tau \le
& \mathcal{C}(\tilde{\epsilon} + |u_0|_{H^2}^2),
\end{split}
\end{equation}
where $\mathcal{C}$ is independent of $u_0$.

Let
\begin{equation} \label{C_1 C_2}
\begin{aligned}
& C_1= C(u_0)\epsilon, &&  C_2=C(u_0) \epsilon ^{-N}, \\
&\tilde{\epsilon}= \epsilon ^{N},  && M=\mathcal{C}(\tilde{\epsilon}
+ |u_0|_{H^2}^2).
\end{aligned}
\end{equation}
It is easy to see that there exists $\epsilon>0$ sufficiently
small such that for any initial data $u_0$ satisfying
$|u_0|_{H^2}^2 \le \epsilon^N$ we have
\begin{equation}
\begin{split}
\alpha B_1  \ge C_1(\mathcal{A}(0)+ C_2M + 4\mathcal{C}).
\end{split}
\end{equation}
Then by the local well-posedness, we know that there exists $T^{\ast}>0$ such
that
\begin{equation} \label{A(t) est. 3}
\begin{split}
\alpha B_1  \ge C_1 \mathcal{A}(t), \quad \text{for $t<T^{\ast}$.}
\end{split}
\end{equation}
 We claim that $T^\ast \ge
\tilde{\epsilon}$. Otherwise, by \eqref{H^2 energy ineq}, \eqref{C_1
C_2} and \eqref{A(t) est. 3}, we have
\begin{equation}
\begin{split}
\frac{\mathrm{d} \mathcal{A}(t)}{\mathrm{d} t} \le C_2
\mathcal{A}(t) \quad \forall \: t \in [0,\, T^\ast],
\end{split}
\end{equation}
then by \eqref{A(t) est. 2}
\begin{equation}
\begin{split}
\mathcal{A}(T^\ast) - \mathcal{A}(0) \le  C_2\int_0^{T^\ast}
\mathcal{A}(\tau) \, \mathrm{d} \tau \le & C_2\int_0^{\tilde{\epsilon}} \mathcal{A}(\tau) \, \mathrm{d} \tau \\
\le & C_2 M,
\end{split}
\end{equation}
which leads to the following contradiction to the definition of
$T^\ast$:
\begin{equation} \label{contradiction}
\begin{split}
\alpha B_1  > C_1 \mathcal{A}(T^\ast).
\end{split}
\end{equation}

We claim now that $T^\ast=\infty$. Otherwise, by \eqref{A(t) est. 2}
$\exists \: t^\ast \in [T^\ast - \frac{\tilde{\epsilon}}{2},\,
T^\ast]$, such that
\begin{equation}
\begin{split}
\mathcal{A}(t^\ast) \le 4\mathcal{C}.
\end{split}
\end{equation}
We also know
\begin{equation}
\begin{split}
\mathcal{A}(T^\ast) - \mathcal{A}(t^\ast)  \le C_2 M.
\end{split}
\end{equation}
Thus
\begin{equation}
\begin{split}
\mathcal{A}(T^\ast) \le 4\mathcal{C} + C_2 M.
\end{split}
\end{equation}
Again, we are led to the contradiction \eqref{contradiction}.

Since $T^\ast = \infty$, then
\begin{equation}\label{111}
\begin{split}
\alpha B_1  \ge C_1 \mathcal{A}(t) = C_1(|u(t)|_{H^2}^2 + 1) \quad
\forall \: t \ge 0,
\end{split}
\end{equation}
which implies the uniform $H^2$ bound of the solution.

Finally, Theorem \ref{Global well-posedness} follows from Proposition
\ref{local existence} and \eqref{111}.
\ep

\subsection{Proof of Theorem~\ref{existence in interpolation spaces}}\label{6.4}
We first give a lemma on the existence of solutions to the following
Cauchy problem:
\begin{equation}  \label{nonhom linear eqn.}
\begin{aligned}
&\frac{\mathrm{d} u}{\mathrm{d} t} = Au + f(t), \\
&u(0)= u_0,
\end{aligned}
\end{equation}
where $A$ is the infinitesimal generator of an analytic semigroup in
$X$ with domain $D(A)$.

\bl  \label{Lunardi's Lemma}

Let $\omega_A = \sup\{\,\mathrm{Re} \, \lambda \: | \: \lambda \in
\sigma(A) \,\} <0$, $f\in C_b([0,\, \infty); \, D_A(\theta))$, and
$u_0 \in D_A(\theta+1)$, where $\sigma(A)$ is the spectral set of
$A$, $D_A(\theta)$ and $D_A(\theta+1)$ are as defined in \eqref{Def
D_A} with some $0 < \theta < 1$. Then there is a unique solution of
\eqref{nonhom linear eqn.} which belongs to $C_b([0,\,\infty);\,
D_A(\theta+1)) \cap C_b^1([0,\,\infty);\, D_A(\theta))$, and there exists a constant $C$ independent of $f$ and $u_0$, such that
\begin{equation}
\begin{aligned}
\| u \|_{C_b([0,\,\infty);\, D_A(\theta+1))} &+ \| u'
\|_{C_b([0,\,\infty);\, D_A(\theta))} \\
& \le C (\|f\|_{C_b([0,\,\infty);\, D_A(\theta))} + \|u_0
\|_{D_A(\theta+1)}).
\end{aligned}
\end{equation}

\el

This lemma is a direct consequence of sections 4.3 and 4.4 of Lunardi
\cite{Lun95}.

We first show that Theorem~\ref{existence in interpolation spaces} is true when $T > T_c$. In this case,
from \eqref{PES} we see that $\omega_{L_T} = \sup\{\, \lambda \: |
\: \lambda \in \sigma(L_T) \,\} <0$. Now for any given $v \in B(0,1)
\subset C_b([0,\,\infty);\, D_{L_T}(\theta+1))$, we consider the
following linear equation:
\begin{equation} \label{Linearized CHE}
\begin{aligned}
&\frac{\mathrm{d} u}{\mathrm{d} t}=L_T u + G(v,T),\\
& u(0)=u_0.
\end{aligned}
\end{equation}

With our choice of $\theta$, the space $D_{L_T}(\theta )$ forms an
algebra according to Lemma \ref{crucial lemma}. Then it is easy to
see that $G(v, T) \in C_b([0,\,\infty);\, D_{L_T}(\theta))$. For
example, the term $v(t)^2\Delta ^2 v(t)$ in $G(v,T)$ can be
estimated as
\begin{equation*}
\begin{aligned}
\|v(t)^2\Delta ^2 v(t) \|_{D_{L_T}(\theta)} & \le
\|v(t)\|_{D_{L_T}(\theta)}
\|v(t)\|_{D_{L_T}(\theta)} \|\Delta^2 v(t)\|_{D_{L_T}(\theta)} \\
& \le \|v(t)\|_{D_{L_T}(\theta)}^2 \|v(t)\|_{D_{L_T}(\theta+1)} \\
& \le \|v(t)\|_{D_{L_T}(\theta+1)}^3, \qquad \forall \: t \ge 0.
\end{aligned}
\end{equation*}
So $v^2 \Delta^2 v \in C_b([0,\,\infty);\, D_{L_T}(\theta))$.
Applying similar estimates to other terms in $G(v, T)$ we obtain the
following:
\begin{equation*}
\begin{aligned}
\|G(v,T) \|_{C_b ([0,\,\infty);\, D_{L_T}(\theta))} \le & C_1
\|v\|^2_{C_b([0,\,\infty);\, D_{L_T}(\theta+1))} +
o(\|v\|^2_{C_b([0,\,\infty);\, D_{L_T}(\theta+1))}),
\end{aligned}
\end{equation*}
where $C_1$ is independent of $v$.

Now, by Lemma \ref{Lunardi's Lemma}, \eqref{Linearized CHE} has a
unique solution $u$ in $C_b([0,\,\infty);\, D_{L_T}(\theta+1)) \cap
C_b^1([0,\,\infty);\, D_{L_T}(\theta))$, which satisfies
\begin{equation} \label{est.for u}
\begin{aligned}
\| u \|_{C_b([0,\,\infty);\, D_{L_T}(\theta+1))} \le & C
\left(\|G(v,T)\|_{C_b([0,\,\infty);\, D_{L_T}(\theta))} + \|u_0
\|_{D_{L_T}(\theta+1)}\right) \\
\le & C_2 \left(\|v\|^2_{C_b([0,\,\infty);\, D_{L_T}(\theta+1))} +
\|u_0 \|_{D_{L_T}(\theta+1)} \right).
\end{aligned}
\end{equation}

Let $R=\min\{\frac{1}{4C_2}, 1\}$, $B_1$ be the ball centered at zero in
$C_b([0,\,\infty);\, D_{L_T}(\theta+1))$ with radius $R$ , and $B_2$ be the ball of
radius $R^2$ centered at zero in $D_{L_T}(\theta+1)$. Define a
mapping $\Gamma$ as follows
$$\Gamma:B_1 \times B_2 \rightarrow B_1, \qquad \Gamma(v, u_0)=u,$$
where $u$ is the solution of \eqref{Linearized CHE} with given $v$
and $u_0$. By \eqref{est.for u} and our choice of $R$, $\Gamma$ is
well defined.

Now we will prove that $\Gamma$ is a contraction in the first
variable. For any $v_1, v_2 \in B_1$, let $\Gamma(v_i, u_0)=u_i$,
$i=1,2$. Let $u=u_1-u_2$ and $v=v_1-v_2$. Then $u$ satisfies the
following equation:
\begin{equation}
\begin{aligned}
&\frac{\mathrm{d} u}{\mathrm{d} t} = L_Tu + G(v_1, T) - G(v_2, T), \\
& u(0) = 0.
\end{aligned}
\end{equation}
Again by Lemma \ref{Lunardi's Lemma}, we have
\begin{equation*}
\begin{aligned}
\| u \|_{C_b([0,\,\infty);\, D_{L_T}(\theta+1))} \le & C
\|G(v_1,T)-G(v_2,T)\|_{C_b([0,\,\infty);\, D_{L_T}(\theta))} \\
\le & C_2 \|v\|_{C_b([0,\,\infty);\,
D_{L_T}(\theta+1))}\big(\|v_1\|_{C_b([0,\,\infty);\, D_{L_T}(\theta+1))} \\
 &+\|v_2\|_{C_b([0,\,\infty);\, D_{L_T}(\theta+1))}\big) \\
\le & 2RC_2 \|v\|_{C_b([0,\,\infty);\, D_{L_T}(\theta+1))} \\
\le & \frac{1}{2} \|v\|_{C_b([0,\,\infty);\, D_{L_T}(\theta+1))}.
\end{aligned}
\end{equation*}
Namely,
$$\| \Gamma(v_1,u_0)-\Gamma(v_2,u_0)\|_{C_b([0,\,\infty);\, D_{L_T}(\theta+1))}
\le \frac{1}{2} \|v_1 - v_2 \|_{C_b([0,\,\infty);\,
D_{L_T}(\theta+1))}.$$

From above, we see that given any $u_0 \in B_2$ there is a unique
fixed point $u \in B_1$ such that $\Gamma(u,u_0)=u$. So for any
initial datum $u_0\in B_2 \subset D_{L_T}(\theta+1)$, the equation
\eqref{CHE. recall} admits a unique solution $u\in
C_b([0,\,\infty);\, D_{L_T}(\theta+1))$. It is easy to see that $u$
is also in $C^1_b([0,\,\infty);\, D_{L_T}(\theta))$. The theorem is
proved in this case with $r=R^2$.

For the case when $T\le T_c$, we define
\begin{equation*}
\begin{aligned}
& \widetilde L_T=L_T -(\beta_1(T)+ \delta) \, id,\\
& \widetilde G(\cdot,T) = G(\cdot,T) + (\beta_1(T)+ \delta) \, id,
\end{aligned}
\end{equation*}
where $id$ is the identity map, $\beta_1(T)$ is the largest eigenvalue
of $L_T$, and $\delta$ is some positive number to be chosen below.

By \eqref{PES}, we can choose $\epsilon$ and $\delta$ sufficiently
small such that
\begin{equation} \label{smallness condition}
|\beta_1(T)+ \delta |< R, \qquad \forall \: T \in[T_c - \epsilon,
T_c],
\end{equation}
where $R=\min\{\frac{1}{4C_2}, 1\}$ as before.

Now consider \eqref{Linearized CHE} with $L_T$ and $G$ replaced by $\widetilde L_T$ and $\widetilde G$, respectively.
Note that $\omega_{\widetilde L_T} = \sup\{\, \lambda \: |
\: \lambda \in \sigma(\widetilde L_T) \,\} <0$. Following the same argument as for the case $T> T_c$ with suitable modification and making use of \eqref{smallness condition},  one can show that for any $u_0 \in B(0, R^2)  \subset D_{L_T}(\theta+1)$, there is a unique $u\in
C_b([0,\,\infty);\, D_{L_T}(\theta+1)) \cap C^1_b([0,\,\infty);\, D_{L_T}(\theta))$ such that
\begin{equation*}
\begin{aligned}
\frac{\mathrm{d} u}{\mathrm{d} t}=& \widetilde L_T u + \widetilde G(u,T)  \\
=& L_T u  - (\beta_1(T)+ \delta) u  + G(u,T) + (\beta_1(T)+ \delta) u \\
=& L_T u   + G(u,T) .
\end{aligned}
\end{equation*}
The proof is now complete.

\bibliographystyle{plain}

\end{document}